\documentclass[fleqn,usenatbib]{mnras}
\usepackage{graphicx}
\usepackage[section]{placeins}
\usepackage{yfonts}

\usepackage{mathptmx}
\usepackage{times} 

\usepackage[T1]{fontenc}
\usepackage{amsmath}
\usepackage{amssymb}
\usepackage{tikz}
\usepackage{xcolor}

\begin{document}

\title[Dynamics of planet-crossing asteroids]{Inclination pathways of planet-crossing asteroids}
\author[F. Namouni]{F. Namouni$^{1}$\thanks{E-mail:namouni@oca.eu} \\
$^{1}$Universit\'e C\^ote d'Azur, CNRS, Observatoire de la C\^ote d'Azur, CS 34229, 06304 Nice, France}

\date{Accepted 2021 November 19. Received 2021 November 5; in original form 2021 July 28}
\pubyear{2022}
\maketitle

\begin{abstract}
Long-term statistical simulations of the past evolution of high-inclination Centaurs showed that their orbits tend to be polar with respect to the solar system's invariable plane over a large semimajor axis range in trans-neptunian space.  Here, we lay the analytical foundation of  the study of the inclination pathways of planet-crossing asteroids that explains these findings. We show that the Tisserand relation partitions the inclination-semimajor axis parameter space of the three-body problem  into distinct regions depending on the asteroid's {Tisserand parameter $T$ or equivalently its } orbital inclination $I_\infty$ far from the planet. The Tisserand relation shows that asteroids with $  I_\infty >110^\circ$ ($T<-1$) cannot be injected inside the planet's orbit. Injection onto retrograde orbits and high-inclination prograde orbits occurs  inside the inclination  corridor $45^\circ \leq I_\infty \leq 110^\circ$ ($-1\leq T\leq 2$). Inclination dispersion across the inclination pathway for moderate and high inclinations is explained by the secular perturbations from the planet and is smallest for polar orbits. When a planet-crossing asteroid temporarily leaves the inclination pathway, its long-term evolution still depends on its Tisserand parameter as evidenced by its eccentricity dispersion.
Simulations of asteroid orbits using the equations of motion with Neptune as the perturbing planet confirm these results for moderate to high inclinations, forward and backward in time because  the Tisserand relation is time-independent. The Tisserand inclination pathways will provide important constraints on comet delivery from the outer solar system as well as on the possible presence of unknown planets in trans-neptunian space. 
\end{abstract}

\begin{keywords}
celestial mechanics--comets: general--Kuiper belt: general--minor planets, asteroids: general -- Oort Cloud. \\[-5mm]
\end{keywords}

\section{Introduction}
Close encounters of small bodies and planets are one of the major mechanisms that drive  the orbital evolution of early disc planetesimals and current asteroid populations in the solar system such as comets, Centaurs and scattered trans-neptunian objects (TNOs). Planet-crossing asteroids in the outer solar system originated from two older reservoirs. One is the early planetesimal disc a small part of which was scattered beyond Neptune's orbit to the Oort cloud region in the first Gyr or so after planet formation \citep{Levison97,TiscarenoMalhotra03,Emelyanenko05,Disisto07,Brasser12,Nesvorny17,Fernandez18,Vokrouhlicky19,Kaib19}.  The other reservoir is the interstellar medium where the sun captured small bodies present in its birth cluster of stars \citep{Fernandez00,Levison10,Brasser12b,Jilkova16,Hands19,Kaib19,NamouniMorais18b,NamouniMorais20b,Portegies21}. After a few Gyr residence in the outer solar system during which they experience the effects of the sun's Galactic environment, asteroids and comets move back on planet-crossing orbits  towards the inner solar system. 

Centaurs, in particular,  are known to have a chaotic dynamical evolution driven by  resonance capture and close encounters with the planets  \citep{TiscarenoMalhotra03,BaileyMalhotra09,VolkMalhotra13,MoraisNamouni13b,NamouniMorais15,MoraisNamouni16,MoraisNamouni17b,NamouniMorais17,NamouniMorais18c}. 
Recently, long-term time-backward numerical simulations of high-inclination Centaurs have shown that their orbits follow distinct pathways in the inclination-semimajor axis plane on time-scales unrelated to their dynamical time \citep{NamouniMorais18b,NamouniMorais20b} (hereafter Papers I and II). It was found that 19 high-inclination Centaur orbits clustered around polar inclinations beyond Neptune's orbit 4.5\,Gyr in the past.  The inclination clustering is the basis for the identification of the interstellar origin of high-inclination Centaurs because trans-neptunian space was empty of solar system material with polar inclinations with respect to the ecliptic $4.5$\,Gyr in the past.

{Clustering around polar inclinations in trans-neptunian space was first seen in the discovery article of TNO (471325) 2011 KT19 where \cite{Chen16} simulated the time-forward evolution of $10^3$ TNO-clones over 1\,Gyr. No attempt was made to investigate the inclination clustering's dynamical origin  probably because the object is currently on a nearly-polar orbit and the polar structure in the clone simulation could have simply reflected the stability of the TNO's inclination. A first indication of the dynamical origin of the polar clustering was given in the million clone simulation of Jupiter's co-orbital asteroid Ka`epaoka`\=awela (Paper I) where the clone swarm was found to follow the asteroid's Tisserand relation, a result  later confirmed in  \citep{kohne20}.  That region of polar motion  in trans-neptunian space was termed the polar corridor  and identified as a prime location for finding sun-bound asteroids of interstellar origin captured in the early solar system (Paper I). In \citep{NamouniMorais20c} (hereafter Paper III), the Tisserand relation was further linked to the dynamics of high-inclination Centaurs  2016 YB13 and 1999 LE31  and the time-backward evolution of TNO  (471325)  2011 KT19. The clustering of high-inclination Centaurs end-states in the polar corridor was also thought to be an artefact of simulating Centaur motion back in time because of a violation of the second law of thermodynamics \citep{mbbr}. In Paper III, it was explained using  physical principles that no such violation occurred. However, the analytical foundation of the  dynamical origin of the polar clustering, whether forward or backward in time, is  still lacking. 

The Tisserand relation is a time-independent function of the asteroid's eccentricity and inclination, and the semimajor axis ratio of the asteroid and the  planet. It  was derived from the exact conservation of the Jacobi constant of the three-body problem \citep{Jacobi36} to identify returning comets whose orbital elements had changed following their encounters with the planets \citep{Tisserand}.  The Tisserand relation was used to characterize small body encounter geometries with Jupiter \citep{Kresak80, Carusi95}, and to distinguish  the orbital types of comets  (ecliptic, Halley-type, Jupiter-family)  \citep{Levison94} and TNOs whose main perturber is Neptune \citep{Elliot05}. If the Tisserand relation is somehow responsible for the inclination clustering of high-inclination Centaurs in a simulation with four giant planets and can indicate the inclination pathways of such planet-crossing asteroids over 4.5\,Gyr, then this result must stem from the three-body problem where the consequences of the conservation of the Tisserand relation can be derived analytically, and the dynamical origin of the polar clustering can be established. This is the aim of this work. 
 
Regardless of their origins and  inclinations, we study analytically and numerically the long-term evolution of Centaurs whose main distinguishing feature we retain is their crossing of the planets' orbits. In effect,  whether they originate in the planetesimal disc or the interstellar medium, Centaurs were drawn back to the planets' domain from trans-neptunian space through the process of close encounters with the planets. 

 In section 2, we recall the derivation of the Tisserand relation in the circular restricted three-body problem and apply it to planet-crossing asteroids.  In section 3, we show analytically  how the Tisserand relation defines a unique portrait of the planet's influence on astroids crossing its orbit through specific inclination pathways that depend on the Tisserand parameter. The Tisserand relation of planet-crossing asteroids partitions parameter space into four dynamical regions. Asteroid injection inside the planet's orbit onto retrograde and high-inclination prograde orbits occurs only in one region that is similar to and contains the polar corridor.  In section 4, we determine analytically the secular dispersion of eccentricity and inclination across the inclination pathway caused by the planet's secular perturbations through the Kozai-Lidov potential.  This helps us confirm the precise agreement of the  analytical theory with the numerical simulations and show how the Tisserand parameter still constrains the asteroid's motion even after it temporarily leaves its Tisserand inclination pathway. In section 5, we use high-resolution statistical simulations  on Gyr time-scales to confirm that the Tisserand inclination pathways  describe accurately the long-term  dynamical evolution of planet-crossing asteroids forward as well as backward in time because the Tisserand relation and the Kozai-Lidov potential are time-independent.  We also show using asteroid orbits with maximal eccentricity that the partitioning of the inclination semimajor axis parameter space by the Tisserand relation is that which corresponds to encounters with the planet at perihelion or aphelion regardless of whether the asteroid's initial perihelion or aphelion is different from the planet's semimajor axis. In section 6, we illustrate asteroid motion in the Tisserand inclination pathways. In section 7, we discuss our findings and their implications for high-inclination Centaurs, comet delivery from the Oort cloud, extreme TNOs, and the presence of unknown planets in trans-neptunian space. }

\vspace{-3mm}

\section{The Tisserand relation of a planet-crossing asteroid}
The equations of motion of the circular restricted three-body problem in the inertial centre-of-mass frame are written as:
\begin{equation}
\ddot{x}={\nabla} GM_\odot |{x}-{x}_\odot|^{-1} +{\nabla} GM_p|{x}-{x}_p|^{-1}\equiv{\nabla} (R_\odot+R_p\nonumber)
\end{equation}
where $M_\odot$ and $M_p$ are the masses of the star and the planet ($M_p\ll M_\odot$), and ${x}$, ${x}_\odot$ and ${x}_p$ are the positions of the asteroid, the sun and the planet  respectively.  As the asteroid is assumed to be massless, the orbits of the sun and planet are circular. Their rotation vector perpendicular to their orbital plane is denoted by ${n}$. The dynamical system has  a constant of motion, the Jacobi constant \citep{Jacobi36}, which is written as \citep{SSDbook}:
\begin{equation}
C_J=-\frac{1}{2}\dot{x}^2+R_\odot+R_p+{n}\cdot ({x}\times \dot{x}).\label{Jacobi}
\end{equation}

\begin{figure*}
{ 
\hspace{-2mm}\includegraphics[width=170mm]{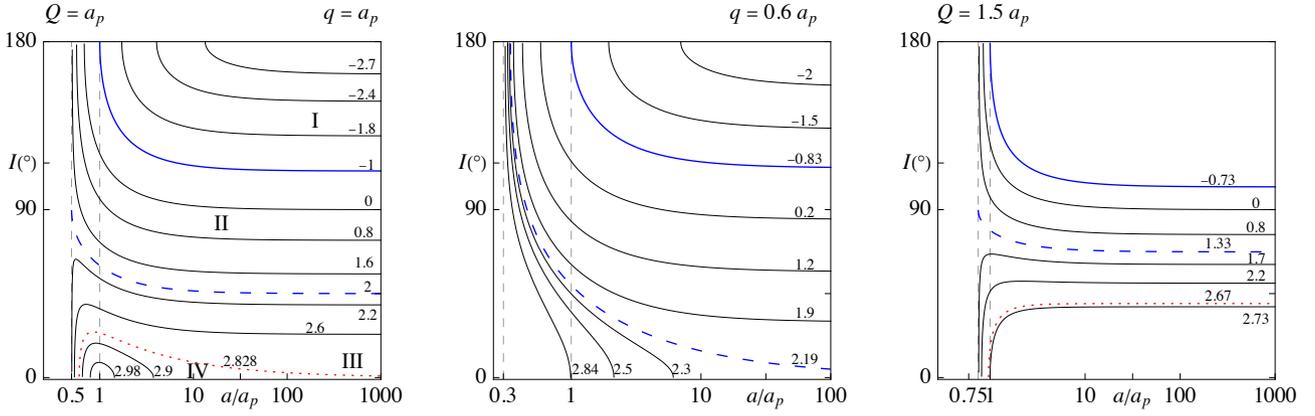}\hspace{10mm} \label{f1}
}
\begin{center}
\caption{Inclination pathways as a function of semimajor axis for various values of the Tisserand parameter. The left-hand panel applies to asteroids initially inside the planet's orbit ($a\leq a_p) $with $Q=a_p$ or outside it with $q=a_p$. The middle panel shows the pathways for an initial $a\geq a_p$ and $q=0.6a_p$. The right-hand panel shows the pathways for an initial $a\leq a_p$ and 
$Q=1.5a_p$. The number above each curve is the Tisserand parameter $T$ (\ref{T0})). The vertical dashed lines denote the smallest semimajor axis accessible to the asteroid at $a_m=\bar qa_p/2$  (left line)  and the planet's location at $a_p$ (right line). 
 The thick solid blue line is the critical Tisserand curve  $T=1-2\sqrt{\bar q(2-\bar q)}$. The dashed blue line is the critical Tisserand curve  $T=2/\bar q$ for $\bar q=1$ and 1.5 and $T=\sqrt{8\bar q}$ for $\bar q=0.6$. The dashed red line is the critical Tisserand curve $T=\sqrt{8\bar q}$ for $\bar q=1$, and $T=4/\bar q$ for $\bar q=1.5$.}\label{f1}
\end{center}
\end{figure*}

After the planet-crossing asteroid moves far away from the planet, its motion around the Sun is purely Keplerian. The first two terms in Eq. (\ref{Jacobi}) then equal the opposite of the asteroid's Keplerian orbital energy  $GM_\odot/2a$, where $a$ is the orbital semimajor axis. The third term vanishes and the fourth is the product of the asteroid's vertical component of its specific angular momentum, $[GM_\odot a (1-e^2)]^{1/2}\cos I$, and the planet's mean motion, $[GM_\odot a_p^{-3}]^{1/2}$ where $e$ and $I$ are the asteroid's orbital eccentricity and inclination respectively and $a_p$ the planet's semimajor axis.  Rearranging the constants $GM_p$ and $a_p$ in $C_J$ leads to the Tisserand relation:
\begin{equation}
T=\frac{a_p}{a}+2 \left[\frac{a(1-e^2)}{a_p}\right]^\frac{1}{2} \cos I. \label{TissRel}
\end{equation}
  The asteroid's encounter with the planet is brief implying that the new post-encounter values of $a$, $e$ and $I$ away from the planet conserve the Tisserand relation \citep{Tisserand}.  Close encounters with a planet tend to conserve the perihelion, $q=a(1-e)$, or the aphelion $Q=a(1+e)$  depending on the encounter's geometry \citep{Duncan87}.  The perihelion and aphelion of a planet-crossing asteroid satisfy  $0\leq q \leq a_p$ and $a_p\leq Q\leq 2 a_p$. The strongest perturbations occur when  $q\sim a_p$ or $Q\sim a_p$ and lead to the fastest evolution of the asteroid's orbit.  
 
 In terms of  the constant $q$ and $Q$, the Tisserand relation may be written as:
\begin{eqnarray}
T&=&\frac{a_p}{a}+2 \left[\frac{q}{a_p}\left(2-\frac{q}{a}\right)\right]^\frac{1}{2} \cos I\ \ \mbox{for}\ a\geq a_p,\label{Tq}\\
T&=&\frac{a_p}{a}+2 \left[\frac{Q}{a_p}\left(2-\frac{Q}{a}\right)\right]^\frac{1}{2} \cos I \ \   \mbox{for}\ a\leq a_p.\label{TQ}
\end{eqnarray}

Three features of the Tisserand relation should  be emphasized. First,  the Tisserand relation is not an exactly conserved quantity along the asteroid's orbit as it is based on taking the limit of the Jacobi constant away from the planet. Secondly, the Tisserand relation of a planet-crossing asteroid given by  (\ref{Tq},\ref{TQ}) is based on the conservation of perihelion (aphelion) distance of a planet-crossing asteroid. Processes such as the secular perturbations of the planet and its mean motion resonances can modify $q$ and $Q$,  so that the asteroid's orbit may no longer satisfy (\ref{Tq},\ref{TQ}). We show in section 4 that the effect of secular perturbations is to cause  inclination and eccentricity dispersions that can be accounted for analytically  as the asteroid temporarily moves away from the orbit-crossing conditions. When mean motion resonances affect the orbit-crossing conditions, they  lead to a radial drift away from the Tisserand inclination pathway (\ref{Tq},\ref{TQ}). In section 5 we show that this occurs only for small inclinations and large Tisserand parameters ($T>2.7$) whereas secular perturbations dominate at moderate to high inclinations. Thirdly, the Tisserand relation defines the inclination pathway of a planet-crossing asteroid but does not indicate its direction of motion along the pathway. Examples of asteroid motion in the inclination pathways forward and backward in time are given in section 6.

\vspace*{-6mm}

\section{Inclination-semimajor axis portrait}
A planet-crossing asteroid with initial orbital elements, $a_0$, $q$ and $I_0$, has a Tisserand parameter:
\begin{equation}
T=\frac{a_p}{a_0}+2 \left[\frac{q}{a_p}\left(2-\frac{q}{a_0}\right)\right]^\frac{1}{2} \cos I_0. \label{T0}
\end{equation}
To derive the pathway the asteroid follows in the inclination-semimajor axis plane, the Tisserand relation (\ref{Tq}) is inverted to obtain: 
\begin{equation}
 I(a,T,q,a_p)=\arccos \left(\left[T-\frac{a_p}{a}\right]{\left[\frac{4q}{a_p}\left(2-\frac{q}{a}\right)\right]^{-\frac{1}{2}}}\right). \label{Tiss}
\end{equation}
{For asteroids inside the planet's orbit with a constant aphelion, the inclination pathway's expression is identical to (\ref{Tiss}) after substituting $Q$ for $q$. The inclination pathway $I(a,T,q$\,or\,$Q,a_p)$ is shown in Fig. (\ref{f1}) as a function of $a/a_p$ and various values of the Tisserand parameter for three values of $q$ and $Q$. In the left-hand panel, the asteroid's initial orbit may be outside the planet's orbit with $q=a_p$  ($a\geq a_p$) or inside it with $Q=a_p$ ($a\leq a_p$). In the middle panel, the asteroid is initially outside the planet's orbit with $q=0.6\,a_p$. In the right-hand panel, the asteroid is initially inside the planet's orbit with $Q=1.5\,a_p$.

To avoid cumbersome notation in describing the mathematical properties of (\ref{Tiss}), we define the normalized perihelion or aphelion distance as $\bar q=q/a_p$  or $Q/a_p$ because the expression of $I(a,T,q,a_p)$ is identical to  $I(a,T,Q,a_p)$. From the perihelion's and aphelion's definitions,  $\bar q=q/a_p$ in the range $0\leq \bar q \leq 1$ and  $\bar q=Q/a_p$ in the range  
$1\leq \bar q \leq 2$.}

The inclination pathway $I(a,T,\bar q,a_p)$ of a planet-crossing asteroid is defined only for $a/a_p\geq \bar q /2$  reflecting the fact that eccentricity is $\leq 1$.\footnote{$\bar q$ here stands for $Q/a_p$.} $I(a,T,\bar q,a_p)$ has an asymptote far from the planet given by:
\begin{equation}
 I_\infty(T,\bar q)=\arccos (T/\sqrt{8\bar q}) \label{Iinfty}
 \end{equation}
for  $-\sqrt{8\bar q}\leq T\leq \sqrt{8\bar q}$. The inclination pathway reaches a maximum given by:
\begin{equation} 
a_m(T,\bar q)=\frac{\bar q a_p}{4-\bar qT},\ \ I_m(T,\bar q)=\arccos \left[\bar q^{-2}\sqrt{\bar q(\bar qT-2)}\right] \label{Im}
\end{equation}
for the following values of the Tisserand parameter: $2/\bar q\leq T\leq \bar q^2+2/\bar q$ for $0\leq \bar q \leq 2^{1/3}$ and $2/\bar q\leq T\leq 4/\bar q$ for $2^{1/3}\leq \bar q \leq 2$. {The inclination pathway leads the asteroid's orbit to coplanarity before it is reflected back on the same pathway. The smallest reflection semimajor axis is given as:}
\begin{equation}
{a_p}{a^{-1}_{180^\circ}(T,\bar q)} = T - 2 \bar q^2 + 2 (\bar q^4- T \bar q^2 + 2 \bar q)^\frac{1}{2}\label{a180}
\end{equation}
defined in the ranges $-\sqrt{8\bar q}\leq T\leq \bar q^2+2/\bar q$ for $0\leq \bar q \leq 2^{1/3}$ and $-\sqrt{8 \bar q}\leq T\leq \sqrt{8\bar q}$ for $2^{1/3}\leq \bar q \leq 2$. The second {reflection} semimajor axis is given by:
\begin{equation}
{a_p}{a^{-1}_{0^\circ}(T,\bar q)} = T - 2 \bar q^2 -2 (\bar q^4- T \bar q^2 + 2 \bar q)^\frac{1}{2}\label{a0}
\end{equation}
and exists only in the range $\sqrt{8\bar q}\leq T\leq \bar q^2+2/\bar q$ and $0\leq \bar q \leq 2^{1/3}$. 

In this work, we seek the analytical foundation of the long-term simulation results of high-inclination Centaurs. Such objects feel the strongest perturbations from the planets {and evolve onto Neptune-crossing orbits with $q\sim a_p$ and $Q\sim a_p$ (see Figure 2 of Paper II). We therefore specialize in planet-crossing asteroids with $q=a_p$ and $Q=a_p$.  In section 4, we examine what happens to the inclination pathways when the conditions $q=a_p$ and $Q=a_p$ are no longer satisfied. In section 5, we show using asteroids with maximal eccentricity orbits that the physically-relevant partitioning of the inclination-semimajor axis parameter space by the Tisserand relation is that with $q=a_p$ and $Q=a_p$  regardless of whether the initial perihelion and aphelion distances are different from $a_p$.}

\begin{figure}
{ 
\hspace{5mm}\includegraphics[width=70mm]{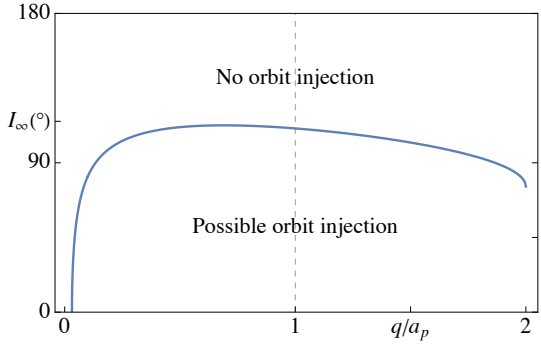}\\[-3mm] \label{f2}
}
\begin{center}
\caption{Minimum inclination at infinity required to inject asteroids inside the planet's orbit as a function of perihelion distance.}\label{f2}
\end{center}
\end{figure}

{For asteroids scattered at perihelion or aphelion, The Tisserand inclination pathway is written as:
\begin{equation}
 I(a,T,a_p)=\arccos \left(\left[T-\frac{a_p}{a}\right]{\left[4\left(2-\frac{a_p}{a}\right)\right]^{-\frac{1}{2}}}\right). \label{Tissq1}
\end{equation}
The Tisserand relation partitions the inclination-semimajor axis plane  into four regions (left-hand panel of Figure 1). 
Region I is that which includes only retrograde motion  (region above the solid blue curve). The corresponding Tisserand parameter $-\sqrt{8}\leq T\leq -1$. In this region, the orbit's inclination $I(a,T,a_p)$ has an asymptote:
\begin{equation}  I_\infty(T)=\arccos (T/\sqrt{8})  \label{Iinftyq1}. \end{equation} 
A far-away asteroid inclined by $I_\infty$ moving towards the planet cannot enter inside its orbit as $a > a_p$. The smallest possible semimajor axis that can be reached is the reflection semimajor axis $a_{180^\circ}$:
\begin{equation}
{a_p}{a^{-1}_{180^\circ}(T)} = T - 2 + 2 ( 3- T)^\frac{1}{2}.\label{a180q1}
\end{equation}
In the vicinity  $a_{180^\circ}$, the incoming asteroid nears retrograde coplanarity ($I=180^\circ$) before it is reflected back on the same Tisserand inclination pathway from where it came. }At the lower boundary of Region I given by $T=-1$, $a_{180^\circ}=a_p$ and  $I_{\infty,1}\simeq 110^\circ$. 

Region II is defined by $-1 \leq T \leq 2$ (region between the solid and dashed blue curves). In this region, asteroids have moderate prograde to polar inclinations far from the planet and may enter inside its orbit. If an asteroid follows the inclination pathway towards its {reflection} semimajor axis  $a_{180^\circ}$, its motion is forced to  become retrograde. {In the vicinity of $a_{180^\circ}$, the incoming asteroid is reflected back on the same Tisserand inclination pathway and may move away from the planet}. No other region allows such evolution. Asteroids starting with aphelia at the planet may move away from its orbit towards the inclination $I_\infty$.  For the lower boundary of Region II, $T=2$ (dashed blue curve) yields $a_{180^\circ}=a_p/2$ and $I_{\infty,2}=45^\circ$. Region II is a high-inclination corridor that  is reminiscent of the polar corridor of high-inclination Centaurs  of Papers I--III that it contains. 

Region III is defined by  $2\leq T\leq \sqrt{8}$ and motion is prograde everywhere (region between the  dashed blue and dotted red curves).  Starting from $I_\infty$ away from the planet, inclination increases along the pathway as the asteroid gets closer to the planet reaching a maximum at (\ref{Im}).
If the asteroid moves closer to the planet, its orbit tends towards the sun-planet orbital plane $I=0^\circ$ at the {reflection} semimajor axis  $a_{0^\circ,1}=a_{180^\circ}$ (\ref{a180q1}) { where it is sent back on the same Tisserand inclination pathway. }The upper boundary of Region III has  {$T=2$ (dashed blue curve)} which  yields $a_m=0.5\,a_p$, and $I_m=90^\circ$.   For the lower boundary of Region III, $T=\sqrt{8}$ (dotted red curve) yields   $I_\infty=0^\circ$, $a_m\simeq 0.85\,a_p$, and $I_m\simeq 24^\circ$.

Region IV is defined by $ \sqrt{8}\leq T\leq 3$ (below the dotted red curve).  {The Tisserand relation indicates that motion may be radially confined to a region defined by $a_{0^\circ,1}\leq a \leq a_{0^\circ,2}=a_{0^\circ}$ (\ref{a0}) where  inclination reaches its maximum at (\ref{Im}). If the asteroid approaches one of the two edges of Region IV, its orbit becomes coplanar with the planet's.  }

The constant perihelion (aphelion) distance may not always be satisfied. If the semimajor axis' change after successive encounters of the asteroid and planet is small, the planet's pull will slowly modify the eccentricity and inclination of the asteroid and may temporarily invalidate the orbit-crossing conditions upon which the inclination pathway is based. This aspect is examined analytically  in the next section and numerically in section 5. The orbit-crossing conditions can also be modified by mean motion resonances with the planet. {The strongest mean motion resonances are close to the planet and will affect motion in Region IV where the semimajor confinement at small inclination and large Tisserand parameters will not occur (section 5).}

When the initial orbit-crossing condition differs from $\bar q=1$, the regions defined by the Tisserand relation are modified except one (see Figure 1). The no-injection Region I exists for all values of $\bar q$. The lower inclination boundary of Region I is given by the equality $a_{\rm 180^\circ}=a_p$ yielding $T=1 - 2 \sqrt{\bar q(2 - \bar q)}$ and :
\begin{equation}
I_\infty^{\rm noinject}=\arccos \left(\frac{1 - 2 \sqrt{\bar q(2 - \bar q)}}{\sqrt{8\bar q}}\right).
\end{equation}
The lower inclination boundary of Region I is shown in Figure (2). $I_\infty^{\rm noinject}\geq 90^\circ$ for most perihelion distances and  reaches a maximum of $112.5^\circ$ at $\bar q=0.689$. Therefore regardless of perihelion distance and the planet's position, planet-crossing asteroids  cannot be injected inside the planet's orbit if their inclination far from the planet $I>I_\infty^{\rm noinject}$. Instead, they are reflected near the semimajor axis $a_{\rm 180^\circ}$ (\ref{a180}) and sent back to the outer solar system. 

\vspace{-4mm}

\section{Secular dispersion across the inclination pathway}
The change to the asteroid's orbital elements following a planet encounter is random and may be large or small. When the semimajor axis change is small over several encounters, the secular perturbations from the planet will build up giving rise to secular periodic changes in the orbital eccentricity and inclination.  This implies that in the secular oscillation cycle the eccentricity will move away from the value that satisfies the orbit-crossing conditions, $q= a_p$ or $Q=a_p$, only to return to it at the cycle's end. This in turn implies that the orbit will temporarily leave  the inclination pathway (\ref{Tissq1})  as the latter is based on the orbit-crossing conditions $q= a_p$ or $Q=a_p$. The asteroid's orbit eventually leaves the secular cycles it experiences at  constant semimajor axis, and returns to the inclination pathway (\ref{Tissq1})  when the amplitude of the semimajor axis change after a planet encounter is large.   The secular oscillations at constant semimajor axis will appear as an inclination dispersion across  the inclination pathway but also as a dispersion in the eccentricity-semimajor axis plane with respect to the orbit-crossing conditions. 
\begin{figure}
\begin{center}
{ 
\hspace{-3mm}\includegraphics[width=85mm]{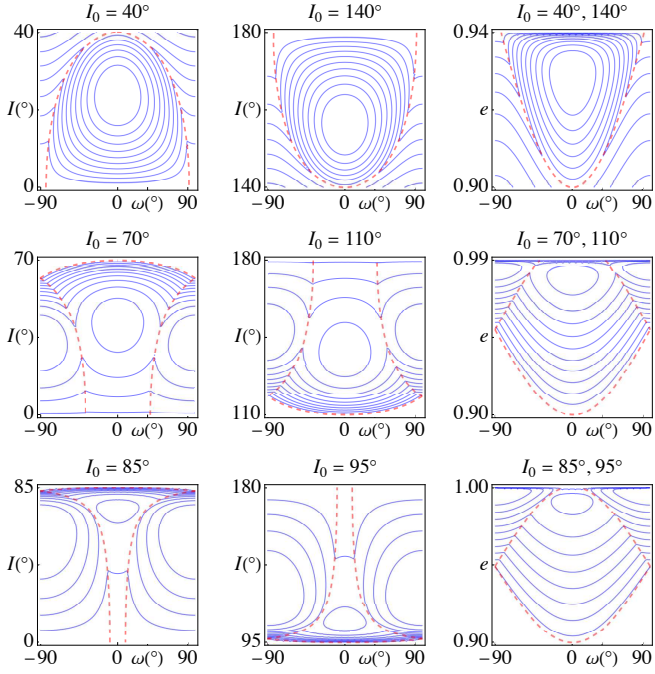}\hspace{10mm}\label{f3}
}
\caption{Level curves (solid blue) of the Kozai-Lidov potential $R_{KL}$ (\ref{RKL}) for $a=10\,a_p$. The dashed red curves are the three-dimensional orbit-crossing conditions. Each row has an inclination-argument of perihelion portrait for an initial inclination, its polar symmetric as well as their common eccentricity-argument of perihelion  portrait. }\label{f3}
\end{center}

\end{figure}

It is possible to find the amplitude of the eccentricity and inclination dispersions analytically. In the circular restricted three-body problem,  the asteroid's orbit  secular evolution is driven by the Kozai-Lidov potential \citep{Kozai62,Lidov62,Quinn90}:
\begin{eqnarray}
R_{KL}&=&\frac{GM_p}{\pi a^2[a_p(1-e^2)]^{1/2}}\int_0^{2\pi}R^{-1/2}{k \, K(k)} r^2 df, \label{RKL}\\
k^2&=& \frac{4Ra_p}{(R+a_p)^2+z^2},\nonumber\\
R^2&=&r^2-z^2,\nonumber\\
z&=&r \sin I \sin(f+\omega),\nonumber
\end{eqnarray}
where $K$ is the complete elliptic integral of the first kind, $r=a (1-e^2)/(1+e \cos f)$ is the radial position of the asteroid and $f$ its true anomaly. The Kozai-Lidov potential conserves the projection of the asteroid's orbital angular momentum  on the rotation vector  ${n}$ of the sun-planet system, $L_z=\sqrt{a(1-e^2)}\cos I$ (section 2). Figure 3 shows the level curves of the Kozai-Lidov potential. In each row, the inclination-argument of pericentre  $(I,\omega)$ level curves are shown for a prograde inclination $I_0$ and its polar symmetric retrograde inclination $180^\circ-I_0$. As the Kozai-Lidov potential has reflection symmetry with respect to exactly polar motion, the two inclination panels are symmetrical with respect to $I=90^\circ$ and the corresponding $(e,\omega)$ level curves are identical. The dashed curves are the three-dimensional orbit-crossing conditions $a_p=a(1-e^2)/(1\pm e \cos \omega)$. The Kozai-Lidov level curves show that for an initially prograde inclination $I_0$ and eccentricity $e_0$ on the orbit-crossing curve at perihelion ($\omega=0^\circ$ and $a_p=a(1-e)$), the inclination travels on a secular cycle that at maximum amplitude makes the asteroid coplanar with the planet. Similarly, a retrograde orbit will travel towards retrograde coplanarity that is a maximum inclination of $180^\circ$ before returning to the initial inclination and eccentricity values.

\begin{figure}
\begin{center}
{ 
\hspace{-2mm}\includegraphics[width=85mm]{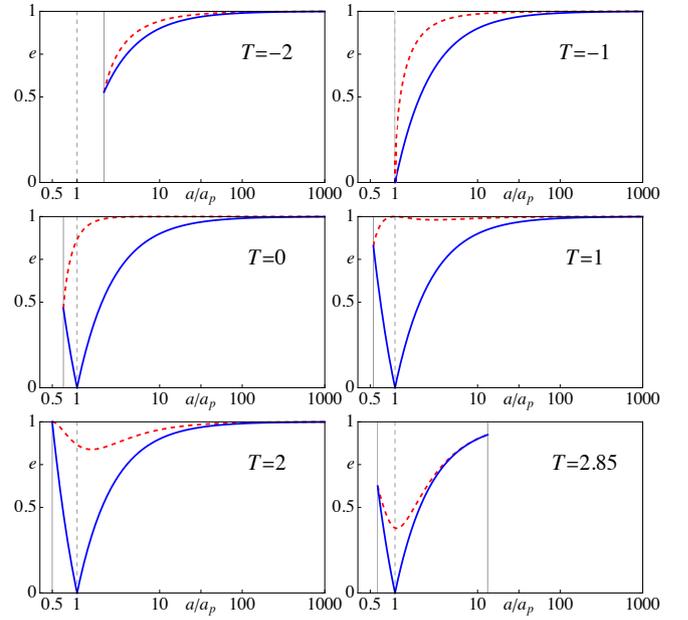}\hspace{10mm}\label{f4}
}
\caption{Eccentricity dispersion amplitude of the Kozai-Lidov oscillations for different values of the Tisserand parameter, $T$. The solid blue curves are the orbit-crossing conditions $q=a_p$ and $Q=a_p$. The dashed red curves are the eccentricity dispersion amplitudes (\ref{emax}). The vertical dashed line is the position of the planet. The vertical solid lines are the locations of the {reflection} semimajor axes  (\ref{a180}) and (\ref{a0}). The latter  appears only in Region IV ($T=2.85$). }
\end{center}

\end{figure}

An asteroid in the inclination pathway  with a Tisserand parameter $T$ whose orbital semimajor axis is no longer strongly perturbed by the planet will leave the pathway with an initial inclination  $I(a,T,a_p)$ (\ref{Tissq1}) { and a vertical angular momentum component given by (\ref{TissRel}) as:
\begin{equation}
L_z=\frac{\sqrt{a_p}}{2}\left(T-\frac{a_p}{a}\right). \label{Lz}
\end{equation}
The inclination dispersion by the Kozai-Lidov potential} will therefore occur in the range $[0^\circ,I(a,T,a_p)]$ for prograde orbits ($I\leq 90^\circ$) and in the range $[I(a,T,a_p),180^\circ]$ for  retrograde orbits ($I\geq 90^\circ$).

Coplanarity (i.e. $I=0^\circ$ or $I=180^\circ$) is achieved for the largest possible eccentricity value. This means that the maximal eccentricity that can be achieved is given by the coplanarity  relation $\pm\sqrt{a(1-e_{\rm max}^2)}= L_z$ because $L_z$ is conserved as soon as the asteroid moves away from the inclination pathway onto the Kozai-Lidov secular cycle. The plus and minus signs indicate prograde and retrograde motion respectively. This relationship and (\ref{Lz}) yield  the eccentricity dispersion amplitude:
\begin{equation}
e_{\rm max}(a,T,a_p)=\left(1-\frac{a_p}{4a}\left[T-\frac{a_p}{a}\right]^2\right)^\frac{1}{2}. \label{emax}
\end{equation}
{Like the vertical angular momentum component,} the eccentricity dispersion does not depend explicitly on the perihelion or aphelion distance but only implicitly through the value of the Tisserand parameter. 

{To the maximal eccentricity corresponds a minimal perihelion distance $q_{\rm min}=a(1-e_{\rm max})$. As the asteroid moves away from the planet $a\rightarrow \infty$ and  $e_{\rm max}\rightarrow  1$ but the minimal perihelion distance is finite and tends to:}
\begin{equation}
q_{\rm min,\infty}=\frac{T^2a_p}{8}.
\end{equation}

The eccentricity dispersion should be viewed alongside the orbit-crossing conditions $q=a_p$ and $Q=a_p$ as these bound the eccentricity from below. Examples  are shown in {Fig. 4}. In the no-injection Region I, eccentricity dispersion is small and confined to the vicinity of the orbit-crossing condition (panel $T=-2$). There is a minimum eccentricity $e_l$ below which the asteroid's orbit cannot go given by $e_{\rm max}(a_{180^\circ},T,a_p)$ at the {reflection} semimajor axis $a_{180^\circ}$. For $T=-2$, $e_l=0.53$. Dispersion increases as Region II is approached ($T=-1$). Inside Regions II and III the maximal eccentricity may reach 1 for small semimajor axes (panels $T=1$ and $2$). The limit eccentricity $e_l$ steadily decreases to zero as $T$ reaches 1 then increases steadily until $T=2$ only to decease afterwards.   Near the lower boundary of Region III, eccentricity dispersion starts to decrease with increasing Tisserand parameter  (panels $T=2$ and $ 2.85$) to ultimately be confined around the planet's semimajor axis.

It should be noted that since the Kozai-Lidov secular potential  (\ref{RKL}) and the Tisserand inclination pathway (\ref{Tissq1}) are both time-independent, time-forward asteroid motion along and across the inclination pathway will be similar to its time-backward motion. This is confirmed using statistical simulations  in the next section.

\vspace{-3mm}

\section{Numerical simulations}

To ascertain the validity of the analytical developments in the previous sections, we perform {long-term} high-resolution statistical simulations of the full equations of motion of the three-body problem with Neptune as the perturbing planet located at $a_p=30.1$\,au on a circular orbit. The asteroid has initial inclination $I_0$ and  semimajor axis $a_0$.  {section 5.1 is devoted to asteroids initially located far outside the planet's orbit with a perihelion distance $q_0=a_p$. In section 5.2 we examine asteroids inside the planet's orbit with an initial aphelion distance $Q_0=a_p$.  In section 5.3, the initial perihelion and aphelion are chosen using  the maximal eccentricity curve (\ref{emax}).} The initial mean longitude relative to Neptune is set to $\lambda=180^\circ$ and the initial argument of perihelion and longitude of ascending node are set to  $\omega=0^\circ$  and $\Omega=0^\circ$. For each initial inclination and semimajor axis, the asteroid's initial orbit is cloned $10^4$ times using a multivariate normal distribution with standard deviations in its equinoctial orbital elements of eccentricity, inclination and mean longitude of $10^{-6}$. For the semimajor axis, a relative standard deviation  $\sigma_a=6\times 10^{-7} a_0$\,(au) was chosen so that  the simulation outcomes of the different initial semimajor axes $a_0$ may be compared to one another particularly in terms of the number of surviving clones. {The clone swarm's orbital elements' deviations from the nominal orbit generate a dispersion in the initial Tisserand parameter, $T_0$, of $\sim 10^{-6}$ to $10^{-5}$ (Table I).}  The orbital elements' standard deviations are similar to the observational error bars of high-inclination Centaurs' orbits (see Paper II). 

The equations of motion were integrated for 1\,Gyr  forward and 1\, Gyr backward in time { for asteroids that encounter the planet at perihelion or aphelion, and  for 2\,Gyr  forward and 2\, Gyr backward in time for asteroids with maximal eccentricities on account of their slower dynamics. The integration uses} the Bulirsch and Stoer algorithm with an error tolerance of $10^{-11}$. Clone orbital evolution stops when one of the following events occurs:  collision with the Sun, collision with Neptune, ejection from the system, reaching the inner 1\,au semimajor axis boundary and reaching the outer boundary at $10^4$\,au.  The choice of the outer boundary in the inner Oort cloud region is motivated by showing the inclination pathway at a large distance from the planet. The simulation does not describe the evolution of asteroids in that region as the gravitational effect of the Galactic environment was not modelled.

{Each set of initial conditions may be characterized by the initial nominal Tisserand parameter, $T_0$ (\ref{T0}), its standard deviation $\sigma_{T_0}$, the inclination at infinity, $I_\infty$ (\ref{Iinftyq1}) and the {reflection}  semimajor axis $a_{180^\circ}$ (\ref{a180q1}). These are given in Table I.  The simulation outcome may be characterized by the mean Tisserand parameter at the simulation's end, $\langle T\rangle$, its standard deviation $\sigma_{T}$ and the inclination standard deviation from the analytical Tisserand pathway inclination (\ref{Tissq1}). The latter is given for two semimajor axis' ranges:  the first is [1:$10^4$\,au] which, as will become apparent below, is equivalent to  [$ a_{180^\circ}$:$10^4$\,au] and gives the global standard deviation. We also use the range   [$3\, a_{180^\circ}$:$10^4$\,au] that indicates the inclination's standard deviation from the analytical pathway inclination away from the planet. The factor $3$ is related to the fact that regardless of inclination, the smallest  {reflection} semimajor axis $a_{180^\circ}= a_p/2$ (section 3). This puts  $3\,a_{180^\circ} $, but not $2\,a_{180^\circ} $, outside the planet's neighbourhood and beyond its strongest perturbations. Another reason for using the second semimajor axis range is that the high-inclination Centaur simulations  (Papers I and II) showed that these objects move away from the planet on polar orbits. The inclination's dispersion observed in those simulations concern semimajor axes larger than 50\,au which fall in the asymptotic part of the inclination pathway (\ref{Tissq1}). The two inclination standard deviations are denoted by $\sigma_{I}$ and $\sigma_{I,a\gg a_p}$. The simulation outcomes are given in Table I.  Unlike our previous works on the origin of high-inclination Centaurs (Papers I and II), we do not quote the collision and ejection fractions of unstable clones because of the presence of the simulation's outer boundary at $10^4$\,au and the absence of three giant planets. We only quote the number of stable clones to indicate the orbit's stability in this work's setting.  }

\begin{table*}
\centering
\caption{Simulation initial conditions and outcome statistics.  For each group of  initial conditions, $T_0$ is the initial nominal Tisserand parameter and $\sigma_{T_0}$ its initial standard deviation. The inclination at infinity (\ref{Iinftyq1}) and the reflection semimajor axis (\ref{a180q1})  are given. The clone number (\#) is quoted for the end of the time-forward and time-backward simulations  denoted by $(+t)$ and $(-t)$ respectively. To the right of each clone number, are the mean Tisserand parameter $\langle T\rangle$  at the corresponding epoch, its standard deviation $\sigma_T$, the global  inclination standard deviation  $\sigma_I$ and  the inclination standard deviation outside the planet's neighbourhood $\sigma_{I, a\gg a_p}$.}
\label{table:1}
\begin{tabular}{ccccccccccccccc}
\hline 
\multicolumn{15}{c}{Asteroids with perihelia at the planet's orbit (Regions I--III): $a_0=300$\,au, $q_0=30.1$\,au, $t=\pm 1$\,Gyr}\\
$I_0$                   &   
$T_0$                                &  
$\sigma_{T_0}$&
$I_\infty$             &  
$a_{180^\circ}$                  & 
Clone \#           & 

$\langle T\rangle$  & 
$\sigma_T$            & 
$\sigma_I$ & 
$\sigma_{I,a\gg a_p}$ &
Clone \#     & 
$\langle T\rangle$  & 
$\sigma_T$     &
$\sigma_I$ & 
$\sigma_{I,a\gg a_p}$\\
($^\circ$)                   &   
                               &  
 $10^{-6}$                              &
($^\circ$)             &  
(au)                 & 
 ($+t$)          & 
      & 
$10^{-4}$ & 
  ($^\circ$)     & 
 ($^\circ$)& 
 ($-t$)&
  & 
 $10^{-4}$ &
($^\circ$) & 
($^\circ$) \\
\hline
140 & $-$2.011287& 9&135& 64.6&4465 & $-$2.0110& 6& 15& 13& 4618& $-$2.0110 & 6&16 &14 \\
130 & $-$1.671526& 8 & 126&46.2 & 3535& $-$1.6710& 6& 15& 14&3450 & $-$1.6710& 6&16  & 13\\
120 & $-$1.277929&  6& 117&35.0 & 2300& $-$1.2772& 6&14 & 11& 2237& $-$1.2772& 6& 13& 10\\
110 & $-$0.842454 & 4&107&27.9 &2743 & $-$0.8421& 9& 8& 5& 2853& $-$0.8421& 9& 8& 6\\
90 & 0.100333       & 3&88& 19.9& 2909& 0.1011& 6& 4& 1& 3143& 0.1011& 7& 4& 2\\
80 & 0.578998       & 5&78& 17.8& 3105& 0.5798& 7&4 & 1& 2626& 0.5798& 7&3 & 1\\
60 & 1.478595       &  9&58&15.5& 694& 1.4797& 7& 11& 10& 764& 1.4796& 7&11 & 9\\
40 & 2.211954        & 11&38& 38.5& 1029& 2.2127& 8& 15& 13& 1080& 2.2127& 8& 15& 14\\
30 & 2.487554      & 13&28& 28.4& 778& 2.4883 & 9&11 & 9& 766& 2.4882 & 9& 10& 9\\
20 &  2.690619      & 13&18& 17.9&375 &  2.6914& 7& 7& 6& 405& 2.6914& 7& 7& 6\\
\hline
\multicolumn{15}{c}{Asteroid with perihelion at the planet's orbit in Region IV: $a_0=50$\,au, $q_0=30.1$\,au, $t= \pm1$\,Gyr}\\
\hline
10 & 2.930815      &2& $-$& 20.7& 4038&  2.9322& 6& $-$ & $-$& 4101& 2.9322& 6&$-$&$-$ \\
\hline
\multicolumn{14}{c}{Asteroids with aphelia at the planet's orbit (Regions I--III): $a_0=20$\,au, $Q_0=30.1$\,au, $t=\pm1$\,Gyr}\\
\hline
170 &  0.119253   & 2&88 &19.9 & 7611& 0.1179& 3& 14& 3&7628 & 0.1179&3 &14 &3 \\
90 & 1.505000      & 2&58 & 15.4& 1360& 1.5035& 5& 11& 9& 1325& 1.5035& 5&11 &10 \\
60 &  2.208562     & 2& 39& 15.1& 3552&  2.2069& 6&16 &13 & 3480&2.2069 & 6&17 & 14\\
20 & 2.827264       & 2& 2& 18.1& 893&  2.8254 & 6& 7& 6& 934&2.8254 &6 &7 & 6\\
\hline
\multicolumn{15}{c}{Asteroid with maximal eccentricity outside the planet's orbit: $a_0=100$\,au, $q_0=5.612$\,au, $t=\pm2$\,Gyr}\\
\hline
0.457 & 1.505000&10& 58&15.4 & 2440& 1.4996& 12& 28& 22& 2464& 1.4996& 12 & 28&22 \\
\hline
\multicolumn{15}{c}{Asteroid with maximal eccentricity inside the planet's orbit: $a_0=20$\,au, $Q_0=38$\,au, $t=\pm2$\,Gyr}\\
\hline
8 & 2.208706    &  3 &39 & 15.1& 1222& 2.1823& 15&25 & 14&1260 & 2.1823& 15 & 25& 14\\

 \hline
\end{tabular}
\end{table*}
\subsection{Asteroids with perihelia at the planet's orbit}
Figure 5 shows the clone distributions at 1\,Gyr forward and at 1\,Gyr backward in time in the inclination-semimajor axis and eccentricity-semimajor axis planes for an initial semimajor axis $a_0=300$\,au, {an initial perihelion $q_0=30.1$\,au}  and decreasing values of the initial inclination $I_0$ from $140^\circ$ to $20^\circ$. {The time-forward dynamics of  such asteroids is representative of the evolution of asteroids and comets moving towards the planet from the inner Oort cloud.} The initial conditions describe asteroids in Regions I--III. 
 The inclination pathways, orbit-crossing conditions $q=a_p$ and $Q=a_p$, eccentricity dispersions $e_{\rm max}$ (\ref{emax}), and the {reflection} semimajor axes  $a_{180^\circ}$ (\ref{a180q1})  are shown.

{Inclinations $I_0=140^\circ,\ 130^\circ$ and $120^\circ$ are located in the no-injection Region I where $a_{180^\circ}>a_p$ or equivalently $T_0<-1$.  The {reflection} semimajor axis varies from $64.6$\,au for $I_0=140^\circ$ to $35.0$\,au for $I_0=120^\circ$.

The Tisserand parameter is conserved with relative standard deviation $\sim 10^{-3}$ for both  time-backward and time-forward simulations (Table I). This deviation amplitude from the initial  Tisserand parameter does not influence the Tisserand inclination pathway (\ref{Tissq1}). } Clones in effect follow the Tisserand pathway for the corresponding initial Tisserand parameter $T_0$ whether the clone swarm's evolution is propagated forward or backward in time. Inclination dispersion occurs precisely between the inclination pathway and retrograde coplanarity as indicated by the Kozai-Lidov secular potential in section 4. A more explicit signature of the Kozai-Lidov potential is shown in the motion of individual clones in section 6.  The inclination standard deviations with respect to the Tisserand pathway inclination generated by secular perturbations are of $\sim 15^\circ$ whether motion is forward or backward in time. For a fixed initial inclination, dispersion decreases with increasing semimajor axis. 

Eccentricity is confined precisely between the perihelion encounter condition, $q=a_p$,  and the maximal eccentricity dispersion $e_{\max}$ (\ref{emax}) obtained from the Kozai-Lidov potential and the Tisserand relation. No clone went inside any of the {reflection} semimajor axes $a_{180^\circ}$.  Orbit instability increases with decreasing inclination as can be seen from the number of clones present at the simulations' end. Although the clone distributions of time-forward and time-back simulations are not identical, the previous results hold precisely for both simulations   because  the Tisserand relation and the Kozai-Lidov potential are time-independent.

\begin{figure*}
{ 
\hspace*{-57mm}\includegraphics[width=300mm]{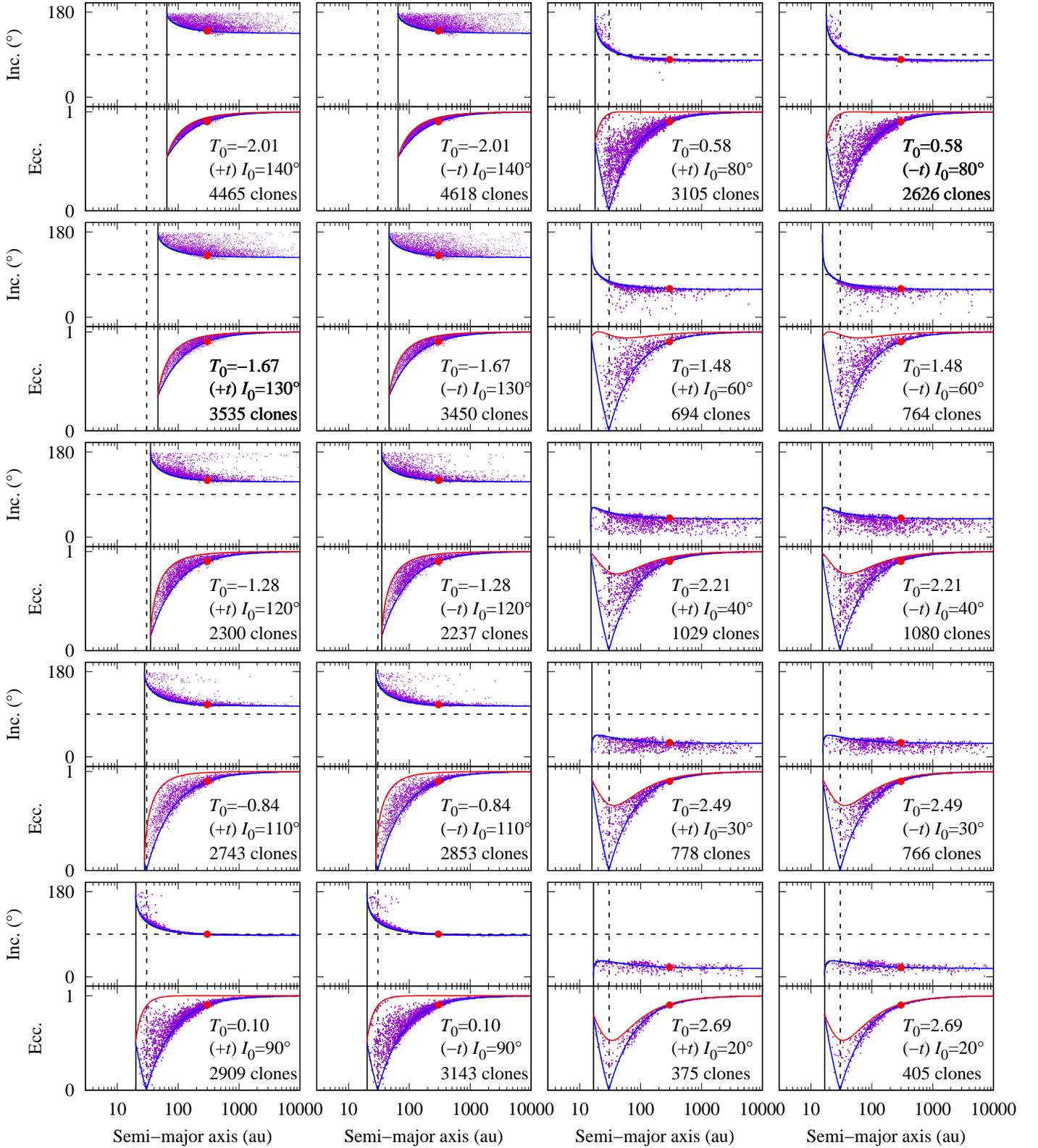}\hspace{10mm} \label{f5}
}
\begin{center}
\caption{Distribution of asteroid clones in the inclination-semimajor axis and eccentricity-semimajor axis planes at $1\,$Gyr in the future and at 1\,Gyr in the past denoted respectively by ($+t$) and  ($-t$) for different initial inclinations $I_0$,  the initial semimajor axis $a_0=300$ au {and the initial perihelion distance $q_0=30.1$\,au. The initial asteroid's position is indicated by the red full circles.} The inclination pathway (\ref{Tissq1}) is the solid blue line in the inclination panels. In the eccentricity panels, the top red curve is the eccentricity dispersion (\ref{emax}) and the bottom blue ones are the orbit-crossing conditions. The vertical dashed line indicates the planet's position and the vertical solid line indicates the {reflection} semimajor axis $a_{180^\circ}$ (\ref{a180q1}). The number of clones present at $\pm1$\, Gyr out of the initial $10^4$ is indicated for each initial condition.}
\end{center}
\end{figure*}

\begin{figure*}
{ 
\hspace*{-57mm}\includegraphics[width=300mm]{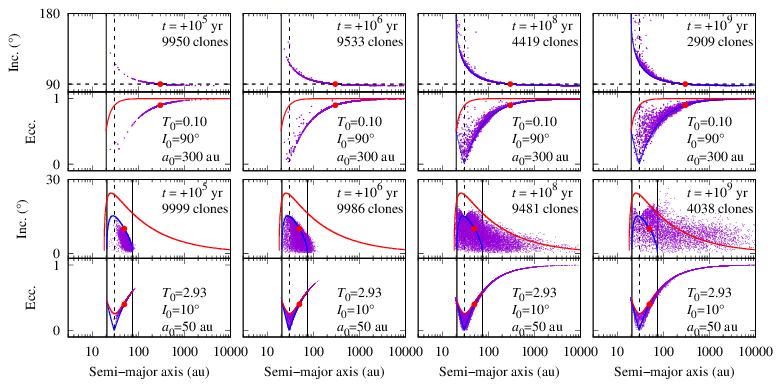}\hspace{10mm} \label{f6}\vspace*{-125mm}
}
\begin{center}
\caption{Snapshots of the clone distribution of two asteroids in Region II ($I_0=90^\circ$,  $a_0=300$\,au, $q_0=30.1$\,au, top row) and  Region IV ($I_0=10^\circ$,  $a_0=50$\,au,  $q_0=30.1$\,au, bottom row)  indicated by the red full circles at epochs 0.1\,Myr, 1\,Myr, 100\,Myr and 1\,Gyr in the future. In the inclination panels, the solid blue line is the inclination pathway (\ref{Tissq1}). In the eccentricity panels, the top red curve is the eccentricity dispersion (\ref{emax}) and the bottom blue ones are the orbit-crossing conditions. The solid red line in the inclination panels of the bottom row is the pathway at the upper boundary of Region IV($T=\sqrt{8}$).  The vertical dashed line indicated the planet's position and the vertical solid lines indicate the {reflection} semimajor axis $a_{0^\circ,1}$ (\ref{a180}) (left) and $a_{0^\circ,2}$ (\ref{a0}) (right), the latter exists only in Region IV. }
\end{center}
\end{figure*}

The precise agreement between analytical theory and simulations  confirms the existence of the no-injection Region I where asteroids heading towards the planet are not allowed to enter inside its orbit but are instead reflected by its gravitational pull near the semimajor axis $a_{180^\circ}$. The presence of Region I will have important consequences on the delivery of comets from the Oort cloud as well as on the signatures of the possible unknown planets in the outer solar system (see section 7).

Inclinations $I_0=110^\circ,\ 90^\circ$, $80^\circ$ and $60^\circ$ fall into Region II, the inclination corridor where incoming asteroids away from the planet can enter its orbit with retrograde and  high-inclination prograde motion. In this region, the Tisserand parameter is conserved  forward and backward in time (Table I). {For the four initial conditions}, inclination dispersion from secular perturbations occurs between the inclination pathway and prograde or retrograde coplanarity depending on the inclination whereas eccentricity dispersion is given accurately by $e_{\rm max}$ (\ref{emax}). This is true whether motion evolves forward or backward in time. 

As the boundary between Regions I and II is crossed with $I_0=110^\circ$, orbit stability increases with inclination, as evidenced by the number of clones, only to decrease as the domain of polar orbits becomes distant and the boundary of Regions II and III is approached. No clones travel inside the {reflection}  semimajor axes $a_{180^\circ}$.

Inclination standard deviations away from the planet $\sigma_{a\gg a_p}\sim 1^\circ$ are minimal for nearly polar orbits. The reason is the reflection symmetry with respect to exactly polar motion of the Kozai-Lidov potential.  

{In the top row of Figure 6, we examine the time evolution of the clones' dispersion as the swarm evolves forward in time from an initial inclination $90^\circ$ namely in the polar corridor of Papers I--III. For the leftmost panels, the Tisserand pathway and the orbit crossing conditions are not shown as dispersion is so small that the clone swarm expands exactly on the inclination pathway for the corresponding  Tisserand parameter, $T_0=0.10$.  At $0.1$\,Myr, a clone already reaches inside the planet's orbit with retrograde motion of orbital elements $a=24.6$\,au, $e=0.23$ and $I= 129^\circ$. At 1\,Myr semimajor axis dispersion  extends from the reflection semimajor axis to the simulation's outer edge [$a_{180^\circ}$, $10^4$\,au]. Epoch $100$\,Myr is somewhat larger than the median lifetime. Clones have had ample time to feel the planet's secular perturbations but the inclination dispersion is moderate and decreases significantly with semimajor axis. At 1\,Gyr, about two-thirds of the clones in the swarm have been lost.  Unstable clones used to follow the Tisserand  inclination pathway just like stable clones do. They become unstable as the ultimate gravitational kick removes  them from the system either by collision or ejection. }

Inclinations $I_0=40^\circ,\ 30^\circ$ and $20^\circ$ fall into Region III where incoming asteroids away from the planet can enter its orbit with prograde motion of moderate inclination. In this region, the Tisserand parameter is conserved forward and backward in time  (Table I). The clones follow the Tisserand inclination pathway in their time-forward and time-backward evolution.  Orbit stability is largest near the boundary of Regions II and III and decreases significantly with the initial inclination. As with the previous initial conditions, no clones travel inside the {reflection} semimajor axes $a_{180^\circ}$. Eccentricity dispersion is given accurately by $e_{\rm max}$. Inclinations are dispersed between prograde coplanar motion and  the theoretical value of the inclination pathway in accordance with the secular evolution driven by the Kozai-Lidov potential.

The examples of Figure 5 thus show that the inclination pathway of the Tisserand relation is followed accurately in Regions I, II and III over a Gyr time-scale forward and backward in time, and that the inclination and eccentricity dispersions are the manifestation of the secular perturbations that build up when the semimajor axis change is small, and take the asteroid temporarily away from the inclination pathway. Inclination dispersion far from the planet is smallest for polar orbits because of the symmetry of the Kozai-Lidov potential with respect to polar motion. Examples of clone motion in the Tisserand inclination pathway are given in section 6.

{Figure 6 (bottom row) shows snapshots of the evolution of the $10^4$-clone swarm of an asteroid with initial semimajor axis $a_0=50$\, au and initial inclination $I_0=10^\circ$ located in Region IV. The pathway parameters are $T_0=2.93$, $a_{0^\circ,1}= 20.7$\,au,  $a_{0^\circ,2}=74.4$\,au and $e_l= 0.46$.  In this region, the Tisserand parameter is conserved forward and backward in time (Table I) but the inclination pathway is not followed and semimajor axis confinement does not occur.} On the time scale of 0.1\,Myr, clone dispersion occurs as predicted by the Tisserand inclination pathway and Kozai-Lidov potential. Eccentricity dispersion is limited by $e_{\rm max}$. At 1\,Myr, clones drift outside the inclination pathway whereas resonances gaps are seen in the swarm distribution. Some clones travel below the {reflection} semimajor axis $a_{180^\circ}$ but semimajor axis diffusion in that direction is minimal at this and subsequent epochs. At 100\,Myr, the inclination amplitude exceeds the inclination pathway and fills almost all of Region IV save the inclination's peak neighbourhood around the coorbital resonance. At 1\, Gyr, the clone swarm reaches its maximum extension and fills out an area in Region III\footnote{This area does not change up to epoch 4.5\,Gyr}. The largest inclination far away from the planet  does not exceed $12^\circ$. 

\begin{figure*}
{ 
\hspace*{-57mm}\includegraphics[width=300mm]{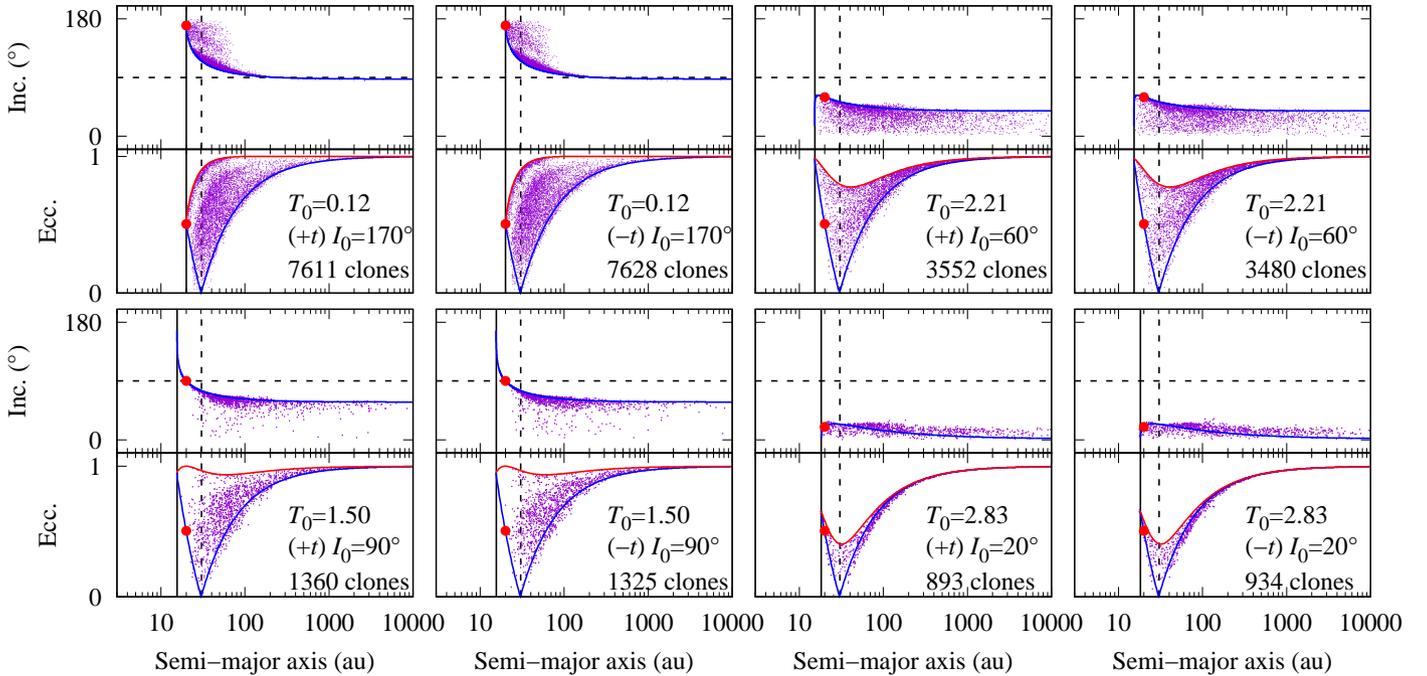}\hspace{10mm} \label{f7}\vspace*{-125mm}
}
\begin{center}
\caption{Same as Fig. 5 but with $a_0=20$ au and $Q_0=30.1$\,au.}
\end{center}\label{f7}
\end{figure*}

{We determined that the deviation from the Tisserand inclination pathway occurs for Tisserand parameters $T > 2.7$ corresponding to an inclination at infinity $I_\infty < 17^\circ$. 
It is interesting to note that inside the inclination pathway region, eccentricity is always bounded by $e_{\rm max}$ regardless of epoch. This is likely related to the fact that the maximal eccentricity (\ref{emax}) is derived from the conservation of the Tisserand parameter and  the conservation of the vertical component of angular momentum. In particular, at no time does  $e_{\rm max}$  require the asteroid to have a nearly constant perihelion distance upon which the Tisserand inclination pathway (\ref{Tissq1}) is based. 

The semimajor axis drift originates in the outer mean motion resonance with Neptune that are strong for prograde nearly-coplanar motion  compared to motion at larger inclinations \citep{NamouniMorais20d}. Since the Tisserand parameter is conserved, finding the analytical Tisserand inclination pathway when mean motion resonances are strong requires finding the relationship between the eccentricity and semimajor axis that replaces that of constant perihelion or aphelion in the Tisserand relation. Evolution in mean motion resonance is known to have constant actions that relate the variations of the orbital elements \citep{SSDbook} (Chapter 6). The equivalent actions for planet-crossing asteroids may lead to the analytical Tisserand pathways in Region IV. This however is beyond the scope of this work which aims to explain the simulation results of high-inclination Centaurs. }

\subsection{Asteroids with aphelia at the planet's orbit}

Figure 7 shows the clone distributions at 1\,Gyr forward and 1\,Gyr backward in time in the inclination-semimajor axis and eccentricity-semimajor axis planes for an initial semimajor axis $a_0=20$\,au, {an initial aphelion $Q_0=30.1$\,au} and four initial inclinations: two in Region II,  $I_0=170^\circ$, and $90^\circ$, and  two in Region III:  $60^\circ$ and  $20^\circ$. {The time-backward dynamics of  such asteroids is representative of the evolution of Centaurs in Paper II.}

In Region II, {the Tisserand parameter is conserved (Table I) and the initially retrograde and polar orbits follow the inclination pathway forward as well as backward in time.} Inclination dispersion is significant in  the planet's neighbourhood  for $I_0=170^\circ$ and occurs between the inclination pathway and retrograde coplanarity. Dispersion declines rapidly as clones move away from the planet onto polar orbits.  Inclination dispersion for $I_0=90^\circ$ occurs between the inclination pathway and prograde coplanarity because the corresponding $I_\infty\sim 60^\circ$ is prograde. Eccentricity dispersion is bounded precisely by $e_{\rm max}$. Orbit stability is largest for $I=170^\circ$. It decreases significantly for $I_0=90^\circ$ as the boundary of Regions II and III is approached and peaks again near $I_0=60^\circ$ ($I_\infty\sim 40^\circ$) after the boundary is crossed. Inclination dispersions caused by the Kozai-Lidov potential are small forward and backward in time, and vary from $3^\circ$  for $I_0=170^\circ$ and $a\gg a_p$ to  $10^\circ$  for $I_0=90^\circ$ and $a\gg a_p$ (see $\sigma_{I,a\gg a_p}$ in Table I).  No clones went below the {reflection} semimajor axis $a_{180^\circ}$.  The clones of the time-backward simulation of  $I_0=170^\circ$ and its  $3^\circ$-inclination dispersion away from the planet are reminiscent of retrograde Centaurs' simulations and their end-states in the polar corridor (Paper II). As explained in section 5.1, the small dispersion with respect to the theoretical pathway inclination occurs because the Kozai-Lidov potential has reflection symmetry with respect to polar motion.

\begin{figure*}
{ 
\hspace*{-57mm}\includegraphics[width=300mm]{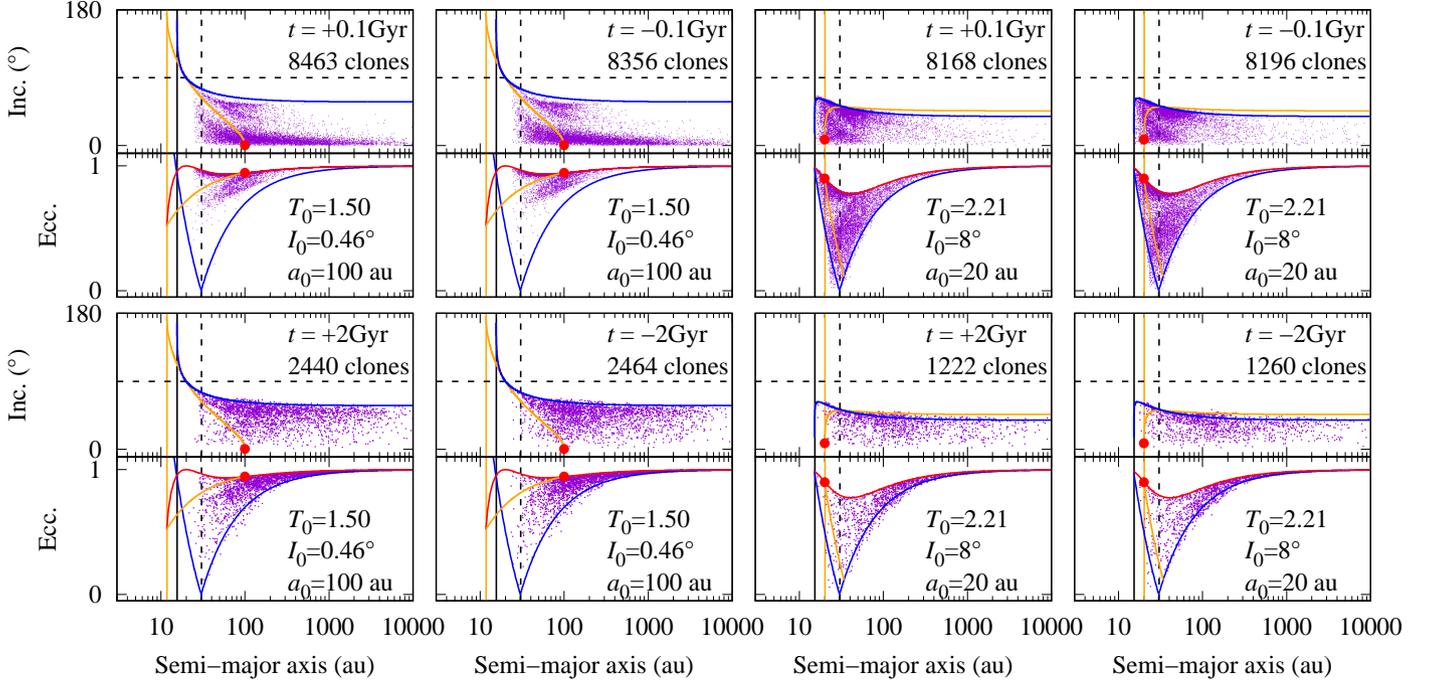}\hspace{10mm} \label{f8}\vspace*{-125mm}
}
\begin{center}
\caption{{Distribution of asteroid clones in the inclination-semimajor axis and eccentricity-semimajor axis planes for two maximal eccentricity asteroids  indicated by the red full circles at four different epochs  ($+0.1$\,Gy and $+2$\,Gyr) in the future and  ($-0.1$\,Gy and $-2$\,Gyr) in the past . The inclination pathway (\ref{Tiss}) with the initial nominal perihelion is the solid orange curve in the inclination panels. The {reflection} semimajor axis $a_{180^\circ} (\ref{a180})$ with the initial nominal perihelion is the vertical orange line. The inclination pathway (\ref{Tissq1}) with is the solid blue curve in the inclination panels. In the eccentricity panels, the top red curve is the eccentricity dispersion (\ref{emax}), the bottom blue curves are the orbit-crossing conditions $q=30.1$\,au and $Q=30.1$\,au, and  the orange curves for $a_0=100$\,au and $a_0=20$\,au  respectively correspond to constant perihelion $q=5.612$\,au and constant aphelion $Q=38$\,au. The black vertical dashed line indicates the planet's position and the black vertical solid line indicates the {reflection} semimajor axis $a_{180^\circ}(\bar q=1)$ (\ref{a180q1}). The number of clones present  is indicated for each epoch.}}
\end{center}\label{f8}
\end{figure*}

In Region III, {the Tisserand parameter is conserved forward and backward in time (Table I),} and eccentricity dispersion is bounded precisely by $e_{\rm max}$ for both initial conditions. Whereas the $I_0=60^\circ$-clone swarm follows precisely the 
inclination pathway,  the $I_0=20^\circ$-swarm tends to a maximum inclination of $18^\circ$ away from the planet instead of  the analytical value of $ I_{\infty}\sim 2^\circ$.  Similarly, whereas no clone goes below the reflection semimajor axis $a_{\rm 0^\circ,1}$ for $I_0=60^\circ$, 3 clones in the time-forward simulation and 2 clones in the time-backward simulation of $I_0=20^\circ$ lie in the semimajor axis range [17.6:18.05]\,au inside $a_{\rm 0^\circ,1}=a_{180^\circ}=18.1$\,au (Table I). This example shows that the lower boundary of Region III is affected by mean motion resonances like Region IV. {This is expected since $T(I_0=20^\circ)=2.827$ is larger than the numerically determined limit of $T=2.7$ where the effect of mean motion resonances is significant. The example of Figure 6 (bottom row) already showed that Region IV clones spill into the lower part of Region III.}  

{The results of the numerical simulations therefore confirm the conservation of the Tisserand parameter in all regions of parameter space whether evolution is propagated forward or backward in time.} The accuracy of the Tisserand inclination pathways (\ref{Tissq1}) forward and backward in time is confirmed {for Tisserand parameters $T<2.7$}. That the eccentricity dispersion, obtained from the combined Tisserand relation  and the Kozai-Lidov potential, is an exact match for the maximal eccentricity in all regions of parameter space is further confirmation of the Tisserand relation's role in the identification of the dynamical pathways of planet-crossing asteroids.

\subsection{Asteroids with maximal eccentricity}
{The previous two sections have demonstrated that the eccentricity of planet-crossing asteroids is always bounded by the orbit-crossing conditions $q=a_p$ and $Q=a_p$ and the maximal eccentricity curve (\ref{emax}) as the perihelion and aphelion distances are modified by the planet's secular perturbations when the semimajor axis change following successive encounters  is small. Understanding motion at maximal eccentricity is particularly useful for three reasons: first, such orbits define the upper boundary of the eccentricity dispersion but also the lower boundary of the inclination dispersion. When the eccentricity nears its maximum, the Kozai-Lidov potential drives the orbit towards near-coplanarity. Therefore whereas their initial inclinations are small their Tisserand parameters can also be small and correspond to motion in Regions I, II and III unlike the low inclination examples in the previous sections 5.1 and 5.2 whose Tisserand parameters approach the maximum value of 3.  Furthermore, since for a given Tisserand parameter $T$, perihelion is smallest at maximal eccentricity for asteroids outside the planet, and aphelion is largest for asteroids outside the planet, studying maximal eccentricity orbits will help us explore the partitioning of inclination-semimajor axis parameter space for orbit crossing conditions that differ from $q=a_p$ and $Q=a_p$. We will show  that the physically meaningful partitioning of parameter space by the Tisserand relation is done with $q=a_p$ and $Q=a_p$   (\ref{Tissq1}) (left-hand panel in Figure 1) regardless of whether the initial perihelion and aphelion distances of the planet-crossing asteroid are different from $a_p$.  Lastly, the orbits of the extreme TNOs with large  eccentricities and low inclinations are similar to those of asteroids near maximal eccentricity that cross the orbits of  possibly unknown planets in the outer solar system.

Figure 8 shows two examples of asteroids with maximal eccentricity located at $100$\,au and $20$\,au interacting with Neptune at 30.1\,au. The $10^4$ clone swarms of each nominal orbit were integrated 2\,Gyr in the future and 2\,Gyr in the past as the dynamics away from the orbit-crossing conditions $q=a_p$  and $Q=a_p$ is slow. 

The outer asteroid has initial nominal inclination $0.457^\circ$, nominal eccentricity $0.94388$ and nominal perihelion 5.612\,au. The inclination pathway (\ref{Tiss}) with its Tisserand parameter $T=1.50$ and its normalized perihelion $\bar q= q/a_p=0.18$ intersects $I=0^\circ$ at $a_{\rm 0^\circ}(1.50,\bar q= 0.18)=100.01$\,au (\ref{a0}) and $I=180^\circ$ at  $a_{\rm 180^\circ}(1.50,\bar q= 0.18)=11.7$\,au (\ref{a180}) (Figure 8, first and second columns). This pathway is representative of those in the middle panel of Figure 1  that seem to transfer asteroids from low to high inclinations deep inside the planet's orbit. The ($T=1.50$,$\bar q=0.18$)-inclination pathway (\ref{Tiss}) and the {reflection} semimajor axis $a_{\rm 180^\circ}(1.50,\bar q= 0.18)$ (\ref{a180}) are shown in Figure 8 in orange along with the location of constant perihelion $q=5.612$\,au.

At $\pm100$\,Myr, the clone swarm disperses away from its original location, ignores the ($T=1.50$, $\bar q=0.18$)-inclination pathway (\ref{Tiss}) and starts to form a clone cloud near the ($T=1.50$, $\bar q=1$)-inclination pathway (\ref{Tissq1}) shown in blue. Similarly, eccentricities start to fill the space between the maximal curve $e_{\rm max}$ and the orbit-crossing conditions $q=a_p$. At $\pm$2\,Gyr, the clones follow the ($T=1.50$,$\bar q=1$)-inclination pathway  with the usual eccentricity and inclination dispersions. For both time directions, the simulations show that the Tisserand parameter is conserved (Table I). The inclination pathway of $T=1.50$ and $\bar q=1$ was already encountered in Figure 5 ($a_0=300$\,au, $I_0=60^\circ$) and in Figure 7 ($a_0=20$\,au, $I_0=90^\circ$). Its inclination at infinity $I_\infty=58^\circ$ (Region II). Its {reflection} radius  $a_{\rm 180^\circ}(\bar q=1)=15.4$\,au (\ref{a180q1}) is shown in Figure 8 as the solid black vertical line. Clones do not travel below $a_{\rm 180^\circ}(\bar q=1)$  in the future or in the past.  
We note that with $\sim2400$ clones at $\pm 2$\,Gyr, the maximal eccentricity asteroid is more stable than those asteroids  with similar Tisserand parameters that encounter the planet at perihelion or aphelion. Their surviving clones at $\pm 1$\,Gyr are $\sim700$ for ($a_0=300$\,au, $I_0=60^\circ$) and $\sim 1300$ for ($a_0=20$\,au, $I_0=90^\circ$).  

The asteroid's evolution is consistent  with the examples of the previous two sections where it was shown that the clone swarms clustered in the inclination pathway (\ref{Tissq1}) of $q=a_p$  regardless of the  secular perturbations that disperses their inclinations between the inclination pathway  and coplanarity and their eccentricities  between the orbit crossing conditions $q=a_p$ and $e_{\rm max}$. More specifically, the simulations in the previous two sections  did not indicate that clone clustering may occur in the long-term near maximal eccentricity. 

This example therefore demonstrates that the low to high inclination transfer pathways in the middle panel of Figure 1 do not represent the true evolution of a planet-crossing asteroid with a perihelion smaller the planet's semimajor axis. The physically relevant parameter space portrait is defined by the Tisserand pathway (\ref{Tissq1})   (left-hand panel of Figure 1).

The inner  asteroid at $a_0=20$\,au in Figure 8 has initial nominal inclination $8^\circ$, nominal eccentricity $0.9$ and  nominal aphelion 38\,au corresponding to an initial Tisserand parameter $T_0=2.21$. The reflection semimajor axis $a_{180^\circ}=19.97$\,au  (\ref{a180}) and the inclination at infinity $I_\infty=46^\circ$ (\ref{Iinfty}) for $\bar q=Q/a_p=1.26$.   The ($T=2.21$,$\bar q=1.26$)-inclination pathway (\ref{Tiss}) and its {reflection} semimajor axis  $a_{180^\circ}$ are shown in orange. In the eccentricity panels, the constant aphelion curve $Q=38$\,au is shown in orange. This example is representative of the region in the right-hand panel of Figure 1 between the dotted red and dashed blue curves.  The Tisserand parameter $T=2.21$ was encountered previously in Figure 5 ($a_0=300$\,au and $I_0=40^\circ$) and in Figure 7 ($a_0=20$\,au and $I_0=60^\circ$) both for $\bar q=1$, and corresponds to motion near the upper boundary of Region III. 

 At $\pm 100$\,Myr, the clone swarm disperses in a similar way to the previous example of $a_0=100$\,au, ignores the ($T=2.21$,$\bar q=1.26$)-inclination pathway (\ref{Tiss}) and starts to form a dense cloud near the ($T=2.21$,$\bar q=1$)-inclination pathway (\ref{Tissq1}) shown in blue in Figure 8. The two pathways are close and have nearby inclinations at infinity ($I_\infty=39^\circ$ for $T=2.21$,$\bar q=1$). However it clear that the swarm follows ($T=2.21$,$\bar q=1$)-inclination pathway as evidenced by the fact that the clones cross the {reflection} semimajor axis $a_{180^\circ}(\bar q=1.26)=19.97$\,au to fill the space below  ($T=2.21$,$\bar q=1$)-inclination pathway but do not cross the {reflection} semimajor axis $a_{\rm 180^\circ}(T=2.21,\bar q=1)=15.14$\, au  (\ref{a180q1}). It should be noted that throughout the future and past evolution, the Tisserand parameter is conserved. 

This example demonstrates that the inclination pathways of asteroids with an initial  $Q>a_p$ are not given by the right-hand panel of Figure 1. The physically-relevant partitioning  of the inclination-semimajor axis parameter space is that of the Tisserand pathway with  $Q=a_p$ (\ref{Tissq1}).  }

\section{Motion in the inclination pathway}
The exact orbit of a planet-crossing asteroid cannot be known on Gyr time-scales. The asteroid's long-term evolution forward and backward in time can be ascertained only statistically. It is still instructive to examine examples of clone motion in order to see the different dynamical processes present as well as the orbits' diversity in the inclination pathway. {The examples will also illustrate how the conserved Tisserand parameter conditions the long-term time-forward and time-backward dynamical evolution of planet-crossing asteroids whether they are in the inclination pathway or leave it temporarily because of secular perturbations. }

Figure 9 contains orbit examples for different initial conditions and different time directions. In order to give an idea about the time residency in parameter space, the orbits are shown in purple for the first 0.1\,Gyr and in green for the subsequent 0.9\,Gyr for which a smaller data sampling is used so as not to overcrowd the Figure. 

\begin{figure*}
{ 
\hspace*{-57mm}\includegraphics[width=300mm]{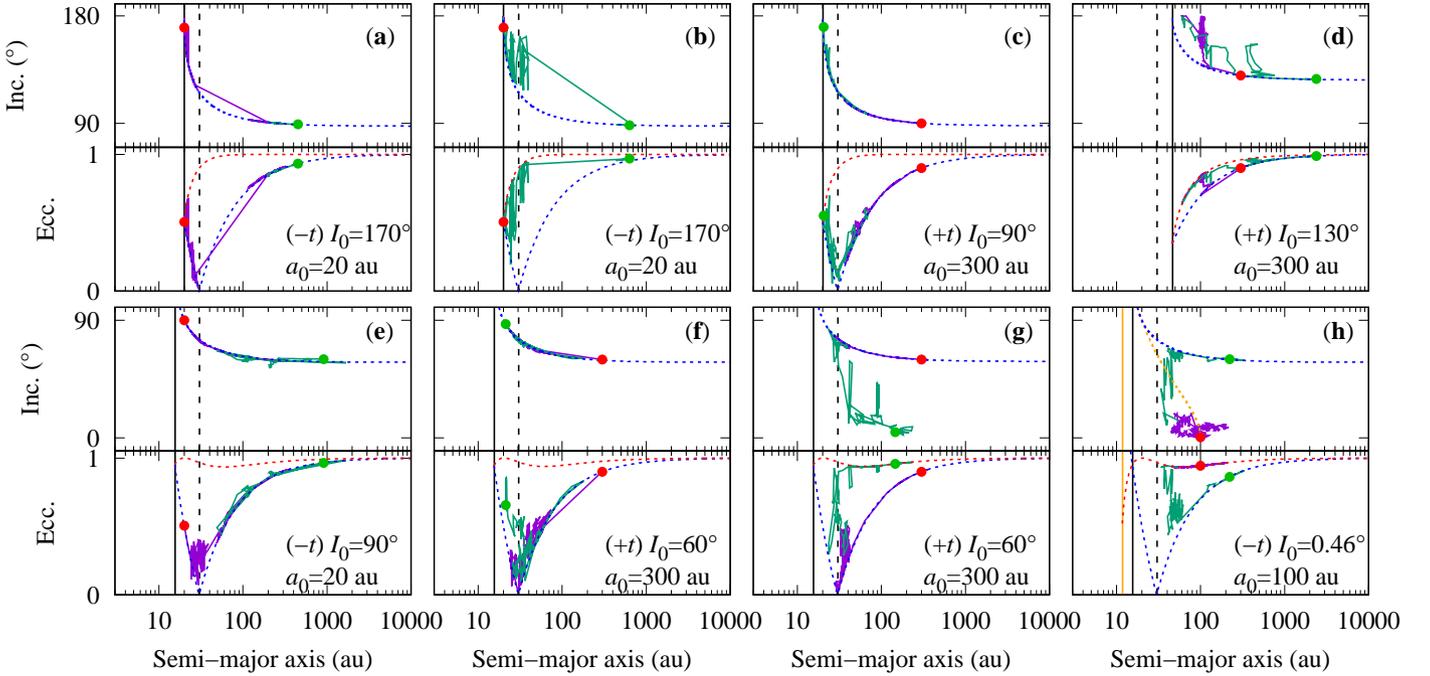}\hspace{10mm} \label{f9}\vspace*{-125mm}
}
\begin{center}
\caption{Examples of clone orbits 1\,Gyr in the past and 1\,Gyr in the future denoted respectively by ($-t$) and  ($+t$) for an initial nominal semimajor axis $a_0$ and initial nominal inclination $I_0$. The red and green full circles are the clone's initial and final positions  respectively. Also shown with dotted lines are the inclination pathway, the orbit-crossing conditions, and the maximal eccentricity dispersion. The black vertical full and dashed lines are the {reflection} semimajor axis $a_{180^\circ}$ and the planet's position respectively. {The orange dotted curve and solid line in Panel (h) correspond to the inclination pathway and {reflection} semimajor axis obtained with the Tisserand parameter and  initial perihelion distance.} The clone's orbit is shown in purple in the first 0.1\,Gyr sampled every $0.1$\,Myr and in green thereafter sampled every $10$\,Myr.}\label{f9}
\end{center}
\end{figure*}

{The first three examples in Panels (a), (b) and (c) illustrate motion in the polar corridor (Papers I--III).} Panel (a) shows an asteroid clone with initial nominal parameters $I_0=170^\circ$,  $a_0=20$\,au {and an aphelion at the planet's orbit} moving backward in time to a polar orbit of  semimajor axis $452$\,au. The clone initially moves along the inclination pathway (\ref{Tissq1}) and across it through secular oscillations seen as the vertical segments indicating  semimajor axis' conservation. These encounters take the clone to the vicinity of the planet and consequently to nearly circular motion. The next planet encounters send the asteroid back on its tracks before expelling it with a single kick away from the planet to 200\,au where it travels back and forth on the inclination pathway. This occurs in the first 0.1\,Gyr. For the subsequent 0.9\,Gyr, the gravitational kicks are small enough so that the clone moves smoothly along the inclination pathway to 452\,au.

Panel (b) shows an asteroid clone with the same initial nominal parameters in Panel (a) moving backward in time to a polar orbit of semimajor axis 633\,au. Unlike the previous example, the clone sticks around its initial orbit for the first 0.1\,Gyr then exhibits high-eccentricity secular oscillations bounded by  $e_{\rm max}$ (\ref{emax}). This results in large inclination oscillations caused by the Kozai-Lidov perturbation and reminiscent of the cloud around the planet's orbit in Fig. 7 ($I_0=170^\circ$). During these cycles, the clone moves steadily outside the planet's neighbourhood until it is removed at high eccentricity  by a single kick onto a polar orbit where it makes its way to its final position.  

Panel (c) shows an incoming polar asteroid clone located at 300\,au {with a perihelion at the planet's orbit}. The clone makes it way forward in time to an orbit inclined by $170^\circ$ of semimajor axis 20\,au similar to the initial orbit of the previous two examples. The clone follows the inclination pathway and the orbit-crossing conditions all the way to its final position passing through a nearly-circular orbit in the vicinity of the planet's location.

Panel (d) shows an  incoming asteroid clone located {initially at 300\,au with a perihelion at the planet's orbit } moving forward in time and {being} reflected by the planet in the no-injection Region I. Its initial nominal inclination is $130^\circ$ and its final semimajor axis and inclination are respectively 2390\,au and $127^\circ$. In the first 0.1\,Gyr, the clone is transferred near a semimajor axis of 106\,au where it  exhibits  Kozai-Lidov oscillations.  After reaching a minimum semimajor axis of 59\,au, the clone is sent back on its way far from the planet.   The eccentricity evolution is strictly confined between the perihelion-encounter condition and $e_{\rm max}$ (\ref{emax}).

Panel (e) shows an initially polar asteroid clone with a nominal semimajor axis of 20\,au and an aphelion at the planet's orbit moving backward in time to a semimajor axis of 913\,au and a prograde inclination of $60^\circ$. The clone travels along the inclination pathway and the orbit-crossing conditions. In the first 0.1\,Gyr, it reaches a maximum semimajor axis 368\,au. During the next 0.9\,Gyr, it achieves a minimum and maximum semimajor axes of 49\,au and 1691\,au respectively before arriving at its final  semimajor axis. 

Panel (f) shows an incoming  asteroid clone located at 300\,au with an initial nominal inclination of $60^\circ$ and {a perihelion at the planet's orbit }moving forward in time to reach a polar orbit at 21\,au at 1\,Gyr {similar to the initial orbit of Panel (e) albeit with a larger eccentricity}. In the first 0.1\,Gyr,  its orbit jumps to 50\,au  and explores motion at small eccentricity in the planet's neighbourhood. In the next 0.9\,Gyr, the clone travels on the inclination pathway to a maximum semimajor axis of 182\,au, {turns around and moves near the orbit-crossing condition towards its final position. }

The last two panels show examples of motion at maximal eccentricity forward and backward in time and further illustrate the conservation of the Tisserand parameter. Panel (g) shows an asteroid clone moving forward in time from an initial nominal inclination of $60^\circ$, semimajor axis of 300\,au and a perihelion at the planet's orbit. The clone follows the orbit-crossing conditions and the inclination pathway in the first 0.1\,Gyr to nearly-circular motion. In the planet's coorbital region, the eccentricity is raised almost to  $e_{\rm max}$ (\ref{emax}). As the clone's motion  is reflected by the planet's gravitational pull and its semi-axis increases, the inclination is lowered to $4^\circ$ at 1\,Gyr corresponding to a semimajor axis of 145\,au whereas the eccentricity follows precisely the eccentricity dispersion curve defined by the asteroid's Tisserand parameter. In time the clone will return to its original Tisserand inclination pathway.

{Panel (h) shows a returning orbit moving backward in time from a low inclination and a Tisserand parameter equal to that in Panel (g). The asteroid clone starts out with  initial nominal inclination $0.46^\circ$, semimajor axis $a_0=100$\,au and perihelion distance $q=5.612$\,au on the maximal eccentricity  curve (\ref{emax}). In the first 0.1\,Gyr it travels back and forth with respect to its initial location on the maximal eccentricity curve. As the clone reaches the planet's vicinity its eccentricity is lowered by the planet's gravitational kicks as well as the secular perturbations (vertical segments) until the asteroid reaches the inclination pathway of its Tisserand parameter (\ref{Tissq1}) that corresponds to the constant perihelion $q=30.1$\,au as explained in section 5.3. The subsequent motion occurs on this pathway where the clone travels to a maximum distance of 332 au and ends up at 221\,au with an inclination $\sim 60^\circ$ at 1\,Gyr near the initial orbit of the example of Panel (g).  The clone's motion is unrelated to the inclination pathway of its initial perihelion shown in Panel (h) as the dotted orange curve. 

 The examples in this section showed how the conserved Tisserand parameter determines the long-term evolution of planet-crossing asteroids forward and backward in time. The examples had  Tisserand parameters $T<2.7$ where the effect of mean motion resonances  is negligible.  They illustrated how the Tisserand inclination pathway is followed, and how maximal eccentricity low inclination planet-crossing orbits may evolve into large inclination orbits. In all our simulations of low inclination asteroid orbits with $T>2.7$ near the lower boundary of Region III and in Region IV, no clone achieved large prograde or retrograde inclinations. }

\vspace*{-1mm}

\section{Discussion}
{In this work, we set out to explain analytically the dynamical origin of the polar clustering found in the time-backward statistical simulations of high inclination Centaurs. To do so, }we reduced the dynamical system where high-inclination Centaurs evolve to its minimum: a planet-crossing asteroid in the three-body problem. In this simplified setting, the planet randomly imparts gravitational kicks to the asteroid's orbit. Their amplitudes 
may be large or small, and may affect the different orbital elements differently. The orbit may also be influenced by mean motion resonances and secular perturbations. All of this makes the long-term orbital evolution of a planet-crossing asteroid  chaotic. In this work, we demonstrated analytically and numerically  that for moderate- to high-inclination asteroids, there is some order within that chaos.  Order is ensured by the Tisserand relation that defines the inclination pathway the planet-crossing asteroid follows (\ref{Tissq1}), and the Kozai-Lidov secular potential that explains the inclination and eccentricity dispersions across the inclination pathway (\ref{RKL},\ref{emax}). {The Tisserand parameter, in particular,  is inscribed on the planet-crossing asteroid's orbit and governs its long-term evolution on Gyr-time-scales whether time flows forwards or backwards.  

The Tisserand inclination pathway depends only on the Tisserand parameter regardless of the asteroid's initial perihelion or aphelion distance.  This was illustrated  for asteroids with maximal eccentricity and hence the smallest perihelion and largest aphelion possible (section 5.3). It was found that the long-term evolution of such asteroids ignores the inclination pathways that correspond to their initial perihelia and aphelia and follows the Tisserand pathways of $q=a_p$ and $Q=a_p$ (\ref{Tissq1}). It can be checked numerically that this result holds for  intermediate perihelion and aphelion distances between the orbit crossing conditions and the maximal eccentricity curve. 

The partitioning of the inclination-semimajor axis parameter space by the Tisserand relation  is reminiscent of the partitioning of motion space by the Jacobi constant and its zero velocity curves \citep{SSDbook}. The latter applies to any kind of motion and gives the regions of permissible and forbidden motion depending on the value of the Jacobi constant. The former applies only to planet-crossing asteroids and gives their inclination pathways.}
The partitioning of parameter space  (left-hand panel of Figure 1) unveiled several novel aspects of the dynamics of planet-crossing asteroids. One  is orbital reflection at the  semimajor axis $a_{180^\circ}$ (\ref{a180q1}). Like the Tisserand inclination pathway (\ref{Tissq1}), the {reflection} semimajor axis does not depend on the initial perihelion or aphelion.

Although its evolution is  chaotic and the asteroid may travel towards or away from the planet (Figure 9), in the long-term it will be reflected by the planet  {at or outside of} $a_{180^\circ}$. Even for small inclinations {and $T>2.7$}, leakage below the reflection radius was minimal and confined to the immediate vicinity of the analytical value $a_{180^\circ}$ indicating that modelling the effect of mean motion resonances will only displace $a_{180^\circ}$  slightly. 

Another novel result is the existence of  the no-injection Region I. 
{For Tisserand parameters $T\leq -1$ (or equivalently inclinations at infinity in the range $110^\circ\leq I_\infty\leq 180^\circ$), $a_{180^\circ}$  is larger the planet's semimajor axis implying that  planet-crossing asteroids will be reflected outside the planet's orbit {and cannot become Centaurs or short-period comets}. This finding will have consequences for the delivery of comets from the spherically shaped Oort cloud as the existence of the no-injection semimajor axis will discriminate between the processes that make comets cross the planets' orbits and the order in which they occur. For instance, Neptune reflects back all comets and asteroids with Tisserand parameters $T_{\rm Neptune}\leq -1$ (or inclinations $I_\infty>110^\circ$). Jupiter the innermost giant planet will do the same and reflect comets with $T_{\rm Jupiter}\leq -1$ but those objects with semimajor axes between Jupiter and Neptune will still enter the giant planets' domain and may become Centaurs and short-period comets. This gives a first limit on the Tisserand parameter of comets driven by close encounters with Jupiter. Setting $a_{180^\circ}(a_{\rm Jupiter})=a_{\rm Neptune}$ where $a_{\rm Jupiter}$ and $a_{\rm Neptune}$ are the planets' semimajor axes, leads to the Tisserand parameter cutoff $T_{\rm Jupiter}\leq -2.53$ (or $I_\infty\geq153^\circ$) below which no comet from the Oort cloud may enter the giant planets' domain. As of 2021, September 3, the JPL Small-Body Database Search Engine\footnote{https://ssd.jpl.nasa.gov} indicates that short-period comets (with periods $<200$\, yr) have Tisserand parameters $T_{\rm Jupiter}>-1.65$ in accordance with the previous cutoff. More interestingly, of the listed 609 long period comets with eccentricities $<1$ (and periods $\geq 200$\,yr), only 5  have Tisserand parameters $T_{\rm Jupiter}<-2.53$.  This suggests that the large majority of the known long period comets will be  or have been able to enter the giant planets' domain in that their semimajor axes could have been or will become smaller than Neptune's. }

The {reflection} semimajor axis $a_{\rm 180^\circ}$ and the no-injection Region I may be used to constrain the existence of unknown planets in the outer solar system \citep{trujillo14,batygin16,malhotra16,volk17,silsbee18,Oldroyd21}. The hypothetical planets are usually assumed to have eccentric and inclined orbits so that they may explain possible orbital alignments of the dynamically-detached TNOs. Inclined planets beyond Neptune would be able to inject Oort cloud comets only with inclinations $0\leq I_\infty\leq 110^\circ$ relative to their orbital plane which, depending on their inclination would in turn  affect the inclination distribution of comets that reach inside Neptune's orbit. Furthermore, the hypothetical planets are assumed to have semimajor axes  $100 \alpha$\,au from the Sun with $\alpha$  of order a few. Therefore TNOs  that cross an outer planet's orbit cannot travel further inwards than the semimajor axis $50\alpha$ (see section 3). The presence of such a TNO population with an inner semimajor axis boundary could be indicative of the planet's position. {In particular,  it would be useful to examine if  the known extreme TNOs form a component of such a population whose dynamics is described by maximal eccentricity motion (section 5.3).} The analysis developed in this paper for a planet on a circular orbit will need to be extended to eccentric planets.

{Going back to the original scope of this article, the polar clustering of high inclination Centaurs can be explained  by the analytical and numerical results in this work. 

In Paper II, a search for the past stable orbits of 19 high inclination Centaurs was performed by integrating the equations of motion with all giant planets and the Galactic tide 4.5\,Gyr in the past using $\sim 2\times 10^7$ clones. The logic of that search was that since Centaurs were believed to have originated from the early planetesimal disc, their orbits are 4.5\,Gyr old. As the gravitational equations of motion are time-reversible, it possible to search statistically for the region of parameter space such objects came from. If no clear dynamical pathways were followed by those Centaurs, then the statistical search would be inconclusive and past orbits would be scattered all over parameter space. Instead, high inclination Centaur orbits clustered mainly in the polar corridor  with small inclination dispersions. In their time-backward evolution high-inclination Centaurs crossed the orbits of the giant planets as they moved outward away from Jupiter's reflection semimajor axis. In the last Gyrs of evolution their clones crossed Neptune's orbit  (Paper II, Figure 2).  From an initial inclination range of [62$^\circ$:173$^\circ$], the 19 high-inclination Centaurs evolved to an average inclination range of  $\sim$[60$^\circ$:90$^\circ$]  with a dispersion $\sim 10^\circ$ in the scattered disc region  at $-4.5$\,Gyr (Paper II Figure 4). Interpreting the latter as a range of  inclinations at infinity (\ref{Iinftyq1}) gives the Tisserand parameter range with respect to Neptune as $0\leq T\leq 1.5$ which  is enlarged to $-0.5\leq T\leq 1.9$ when the inclination dispersion is taken into account. These ranges correspond to Neptune's Region II (Figure I, left-hand panel). The small inclination dispersion comes from the polar symmetry of the Kozai-Lidov potential not only of Neptune but of all giant planets because the relative inclinations of the giant planets' orbital planes are negligible. 

A recurring question about the simulations of high-inclination Centaurs  is whether the inclination pathways they follow in   their time-backward evolution are related to those they follow in a time-forward simulation that starts following their motion far away from the planets as they arrive from the Oort cloud early in solar system history. The simulations in Figures 5 and 7 answer this question. Consider the planet-crossing asteroid in Figure 7 (top row, second panel from left)  located at $a_0=20$\,au and inclined by $I_0=170^\circ$ with a Tisserand parameter $T=0.12$. The asteroid's evolution is propagated backward in time. After 1\,Gyr the clone swarm disperses along the Tisserand inclination pathway away from the planet. This asteroid is representative of Centaurs with retrograde motion such as those examined in Paper II. The swarm's inclination clusters around $88^\circ\pm3^\circ$ (Table I).   In order to determine if such dispersed clones away from the planet can actually follow the same inclination pathway when their motion is reversed in time, we may simulate the time-forward evolution of a clone swarm with an inclination of $90^\circ$ with a semimajor axis far from the planet. This is done in Figure 5 (bottom row, left-hand panel) and Figure 6 (top row) for  $a_0=300$\,au. The clone swarm disperses as predicted on the same Tisserand inclination pathway with  inclination dispersions similar to those of the inner retrograde asteroid. Examples of individual clones of these two simulations are given in Figure 9, panels (a), (b) and (c). Further corresponding examples on either side of the planet with similar Tisserand parameters are given in Figures 7 and 5: ($I_0=90^\circ$, $a_0=20$\,au, negative time) $\leftrightarrow$ ($I_0=60^\circ$, $a_0=300$\,au, positive time), and ($I_0=60^\circ$, $a_0=20$\,au, negative time) $\leftrightarrow$ ($I_0=40^\circ$, $a_0=300$\,au, positive time).    }

{To understand more precisely how the past orbits of high-inclination Centaurs evolve in time, it is necessary to examine their motion in the presence of all the giant planets. This is beyond the scope of this paper but it is possible to identify how the current results derived  in the context of the three-body problem  will be modified.  First, the Centaur's Tisserand parameter with any planet will not be conserved as it is a purely three-body invariant. The evolution of the Tisserand parameters with the giant planets on the dynamical time-scale occurs as the Centaur's clone swarm disperses and moves towards Neptune. The Centaur's inclination pathway will depend on the evolution of its perihelion as it too is no longer conserved  but will secularly move towards Neptune's semimajor axis.  When the clone swarm's mean perihelion has reached Neptune, the Tisserand parameter becomes constant in the scattered disc region and the Tisserand inclination pathways apply. In the inner Oort could region, however, the Galactic environment modifies the Centaur's eccentricity and inclination  and consequently the inclination pathway. The second way this work's results are modified concerns the collision and ejection rates. Centaurs are more stable in the three-body problem than they are in the solar system. The asteroid in Figure 7 with an initial  inclination of $170^\circ$ has a median lifetime $>1$\,Gyr as it still had 7611 clones at $+1$\,Gyr out of the initial $10^4$. In Paper II the two Centaurs with a similar inclination (330759) 2008 SO218 and (434620) 2005 VD have median lifetimes $\sim 2$\,Myr. The presence of all the planets therefore does not change the inclination pathways in the scattered disc but by decreasing the Centaurs' stability it modifies  the determination of the original population's size with respect to a three-body treatment. 
}

The analysis of the agreement between the analytical and numerical results in this work was possible because of the high-resolution statistical simulation that concentrates a large number of clones in a minuscule neighbourhood of the asteroid's orbit. This  approach was initiated with the million clone simulation of  Jupiter's co-orbital asteroid (Paper I). Its importance as illustrated in this work goes beyond the chaotic dynamics of high-inclination Centaurs  to the evolution of the early planetesimal disc. In the early solar system, planets are thought to have  undergone an unstable dynamical phase where Neptune is pushed into the early planetesimal disc whose planetesimals it scatters from the immediate vicinity of its orbit to the Oort cloud \citep{Pfalzner15}. The example in Figure 6 (bottom row) illustrates what happens to a low-inclination asteroid scattered by Neptune.  After an evolution over 1\,Gyr, the $10^4$-clone swarm initially inclined by $10^\circ$ and confined to a range of  $10^{-4}$\,au  produces a disc  that extends from 20\, au to $10^4$\,au (our arbitrary outer simulation boundary  in the absence of Galactic effects).   It is interesting to note the distribution similarity of our Figure 6 with Figure 6 Panels (a) and (b) in \citep{Nesvorny17} where the authors study the dynamics of short period comets from the effect of the planets' instability on a 10\,au-wide disc of a million planetesimals. Despite the inclination amplitude difference and the presence of four planets  in the short-period comet simulation, the inclination similarity likely indicates the presence of inclination pathways the disc planetesimals follow when excited by Neptune's perturbations. Analytical modelling of the effect of mean motion resonances on the Tisserand relation for small inclinations may help uncover the pathways followed by disc planetesimals in the early solar system. 

\section*{Acknowledgments}
I thank the reviewer for constructive comments that helped improve the clarity of the paper. The numerical simulations were done at the M\'esocentre SIGAMM hosted at the Observatoire de la C\^ote d'Azur. 
 \bibliographystyle{mnras}

\vspace*{-3mm}

\section*{Data availability}
The data underlying this article will be shared on reasonable request to the  author.

\vspace*{-5mm}

\bibliography{ms}

\begin{thebibliography}{}
\makeatletter
\relax
\def\mn@urlcharsother{\let\do\@makeother \do\$\do\&\do\#\do\^\do\_\do\%\do\~}
\def\mn@doi{\begingroup\mn@urlcharsother \@ifnextchar [ {\mn@doi@}
  {\mn@doi@[]}}
\def\mn@doi@[#1]#2{\def\@tempa{#1}\ifx\@tempa\@empty \href
  {http://dx.doi.org/#2} {doi:#2}\else \href {http://dx.doi.org/#2} {#1}\fi
  \endgroup}
\def\mn@eprint#1#2{\mn@eprint@#1:#2::\@nil}
\def\mn@eprint@arXiv#1{\href {http://arxiv.org/abs/#1} {{\tt arXiv:#1}}}
\def\mn@eprint@dblp#1{\href {http://dblp.uni-trier.de/rec/bibtex/#1.xml}
  {dblp:#1}}
\def\mn@eprint@#1:#2:#3:#4\@nil{\def\@tempa {#1}\def\@tempb {#2}\def\@tempc
  {#3}\ifx \@tempc \@empty \let \@tempc \@tempb \let \@tempb \@tempa \fi \ifx
  \@tempb \@empty \def\@tempb {arXiv}\fi \@ifundefined
  {mn@eprint@\@tempb}{\@tempb:\@tempc}{\expandafter \expandafter \csname
  mn@eprint@\@tempb\endcsname \expandafter{\@tempc}}}

\bibitem[\protect\citeauthoryear{{Bailey} \& {Malhotra}}{{Bailey} \&
  {Malhotra}}{2009}]{BaileyMalhotra09}
{Bailey} B.~L.,  {Malhotra} R.,  2009, \mn@doi [\icarus]
  {10.1016/j.icarus.2009.03.044}, \href
  {http://adsabs.harvard.edu/abs/2009Icar..203..155B} {203, 155}

\bibitem[\protect\citeauthoryear{{Batygin} \& {Brown}}{{Batygin} \&
  {Brown}}{2016}]{batygin16}
{Batygin} K.,  {Brown} M.~E.,  2016, \mn@doi [\aj]
  {10.3847/0004-6256/151/2/22}, \href
  {https://ui.adsabs.harvard.edu/abs/2016AJ....151...22B} {151, 22}

\bibitem[\protect\citeauthoryear{{Brasser}, {Duncan}, {Levison}, {Schwamb}  \&
  {Brown}}{{Brasser} et~al.}{2012a}]{Brasser12b}
{Brasser} R.,  {Duncan} M.~J.,  {Levison} H.~F.,  {Schwamb} M.~E.,   {Brown}
  M.~E.,  2012a, \mn@doi [\icarus] {10.1016/j.icarus.2011.10.012}, \href
  {https://ui.adsabs.harvard.edu/abs/2012Icar..217....1B} {217, 1}

\bibitem[\protect\citeauthoryear{{Brasser}, {Schwamb}, {Lykawka}  \&
  {Gomes}}{{Brasser} et~al.}{2012b}]{Brasser12}
{Brasser} R.,  {Schwamb} M.~E.,  {Lykawka} P.~S.,   {Gomes} R.~S.,  2012b,
  \mn@doi [\mnras] {10.1111/j.1365-2966.2011.20264.x}, \href
  {http://adsabs.harvard.edu/abs/2012MNRAS.420.3396B} {420, 3396}

\bibitem[\protect\citeauthoryear{{Carusi}, {Kres{\'a}k}  \&
  {Valsecchi}}{{Carusi} et~al.}{1995}]{Carusi95}
{Carusi} A.,  {Kres{\'a}k} {\v{L}}.,   {Valsecchi} G.~B.,  1995, \mn@doi [Earth
  Moon and Planets] {10.1007/BF00671499}, \href
  {https://ui.adsabs.harvard.edu/abs/1995EM&P...68...71C} {68, 71}

\bibitem[\protect\citeauthoryear{{Chen} et~al.,}{{Chen} et~al.}{2016}]{Chen16}
{Chen} Y.-T.,  et~al., 2016, \mn@doi [\apjl] {10.3847/2041-8205/827/2/L24},
  \href {http://adsabs.harvard.edu/abs/2016ApJ...827L..24C} {827, L24}

\bibitem[\protect\citeauthoryear{{Di Sisto} \& {Brunini}}{{Di Sisto} \&
  {Brunini}}{2007}]{Disisto07}
{Di Sisto} R.~P.,  {Brunini} A.,  2007, \mn@doi [\icarus]
  {10.1016/j.icarus.2007.02.012}, \href
  {http://adsabs.harvard.edu/abs/2007Icar..190..224D} {190, 224}

\bibitem[\protect\citeauthoryear{{Duncan}, {Quinn}  \& {Tremaine}}{{Duncan}
  et~al.}{1987}]{Duncan87}
{Duncan} M.,  {Quinn} T.,   {Tremaine} S.,  1987, \mn@doi [\aj]
  {10.1086/114571}, \href
  {https://ui.adsabs.harvard.edu/abs/1987AJ.....94.1330D} {94, 1330}

\bibitem[\protect\citeauthoryear{{Elliot} et~al.,}{{Elliot}
  et~al.}{2005}]{Elliot05}
{Elliot} J.~L.,  et~al., 2005, \mn@doi [\aj] {10.1086/427395}, \href
  {https://ui.adsabs.harvard.edu/abs/2005AJ....129.1117E} {129, 1117}

\bibitem[\protect\citeauthoryear{{Emel'yanenko}, {Asher}  \&
  {Bailey}}{{Emel'yanenko} et~al.}{2005}]{Emelyanenko05}
{Emel'yanenko} V.~V.,  {Asher} D.~J.,   {Bailey} M.~E.,  2005, \mn@doi [\mnras]
  {10.1111/j.1365-2966.2005.09267.x}, \href
  {http://adsabs.harvard.edu/abs/2005MNRAS.361.1345E} {361, 1345}

\bibitem[\protect\citeauthoryear{{Fern{\'a}ndez} \& {Brunini}}{{Fern{\'a}ndez}
  \& {Brunini}}{2000}]{Fernandez00}
{Fern{\'a}ndez} J.~A.,  {Brunini} A.,  2000, \mn@doi [\icarus]
  {10.1006/icar.2000.6348}, \href
  {https://ui.adsabs.harvard.edu/abs/2000Icar..145..580F} {145, 580}

\bibitem[\protect\citeauthoryear{{Fern{\'a}ndez}, {Helal}  \&
  {Gallardo}}{{Fern{\'a}ndez} et~al.}{2018}]{Fernandez18}
{Fern{\'a}ndez} J.~A.,  {Helal} M.,   {Gallardo} T.,  2018, \mn@doi [\planss]
  {10.1016/j.pss.2018.05.013}, \href
  {http://adsabs.harvard.edu/abs/2018P%26SS..158....6F} {158, 6}

\bibitem[\protect\citeauthoryear{{Hands}, {Dehnen}, {Gration}, {Stadel}  \&
  {Moore}}{{Hands} et~al.}{2019}]{Hands19}
{Hands} T.~O.,  {Dehnen} W.,  {Gration} A.,  {Stadel} J.,   {Moore} B.,  2019,
  \mn@doi [\mnras] {10.1093/mnras/stz1069}, \href
  {https://ui.adsabs.harvard.edu/abs/2019MNRAS.tmp.1064H} {419, 1064}

\bibitem[\protect\citeauthoryear{{Jacobi}}{{Jacobi}}{1836}]{Jacobi36}
{Jacobi} C. G.~J.,  1836, Comptes Rendus de l'Acad{\'e}mie des Sciences de
  Paris., 3, 59

\bibitem[\protect\citeauthoryear{{J{\'\i}lkov{\'a}}, {Hamers}, {Hammer}  \&
  {Portegies Zwart}}{{J{\'\i}lkov{\'a}} et~al.}{2016}]{Jilkova16}
{J{\'\i}lkov{\'a}} L.,  {Hamers} A.~S.,  {Hammer} M.,   {Portegies Zwart} S.,
  2016, \mn@doi [\mnras] {10.1093/mnras/stw264}, \href
  {https://ui.adsabs.harvard.edu/abs/2016MNRAS.457.4218J} {457, 4218}

\bibitem[\protect\citeauthoryear{{Kaib} et~al.,}{{Kaib} et~al.}{2019}]{Kaib19}
{Kaib} N.~A.,  et~al., 2019, \mn@doi [\aj] {10.3847/1538-3881/ab2383}, \href
  {https://ui.adsabs.harvard.edu/abs/2019AJ....158...43K} {158, 43}

\bibitem[\protect\citeauthoryear{{K{\"o}hne} \& {Batygin}}{{K{\"o}hne} \&
  {Batygin}}{2020}]{kohne20}
{K{\"o}hne} T.,  {Batygin} K.,  2020, \mn@doi [Celestial Mechanics and
  Dynamical Astronomy] {10.1007/s10569-020-09985-1}, \href
  {https://ui.adsabs.harvard.edu/abs/2020CeMDA.132...44K} {132, 44}

\bibitem[\protect\citeauthoryear{{Kozai}}{{Kozai}}{1962}]{Kozai62}
{Kozai} Y.,  1962, \mn@doi [\aj] {10.1086/108790}, \href
  {http://adsabs.harvard.edu/abs/1962AJ.....67..591K} {67, 591}

\bibitem[\protect\citeauthoryear{{Kres\'ak}}{{Kres\'ak}}{1980}]{Kresak80}
{Kres\'ak} {\v L}.,  1980, \mn@doi [Moon and Planets] {10.1007/BF00896869},
  \href {https://ui.adsabs.harvard.edu/abs/1980M&P....22...83K} {22, 83}

\bibitem[\protect\citeauthoryear{{Levison} \& {Duncan}}{{Levison} \&
  {Duncan}}{1994}]{Levison94}
{Levison} H.~F.,  {Duncan} M.~J.,  1994, \mn@doi [\icarus]
  {10.1006/icar.1994.1039}, \href
  {https://ui.adsabs.harvard.edu/abs/1994Icar..108...18L} {108, 18}

\bibitem[\protect\citeauthoryear{{Levison} \& {Duncan}}{{Levison} \&
  {Duncan}}{1997}]{Levison97}
{Levison} H.~F.,  {Duncan} M.~J.,  1997, \mn@doi [\icarus]
  {10.1006/icar.1996.5637}, \href
  {http://adsabs.harvard.edu/abs/1997Icar..127...13L} {127, 13}

\bibitem[\protect\citeauthoryear{{Levison}, {Duncan}, {Brasser}  \&
  {Kaufmann}}{{Levison} et~al.}{2010}]{Levison10}
{Levison} H.~F.,  {Duncan} M.~J.,  {Brasser} R.,   {Kaufmann} D.~E.,  2010,
  \mn@doi [Science] {10.1126/science.1187535}, \href
  {http://adsabs.harvard.edu/abs/2010Sci...329..187L} {329, 187}

\bibitem[\protect\citeauthoryear{{Lidov}}{{Lidov}}{1962}]{Lidov62}
{Lidov} M.~L.,  1962, \mn@doi [\planss] {10.1016/0032-0633(62)90129-0}, \href
  {http://adsabs.harvard.edu/abs/1962P%26SS....9..719L} {9, 719}

\bibitem[\protect\citeauthoryear{{Malhotra}, {Volk}  \& {Wang}}{{Malhotra}
  et~al.}{2016}]{malhotra16}
{Malhotra} R.,  {Volk} K.,   {Wang} X.,  2016, \mn@doi [\apjl]
  {10.3847/2041-8205/824/2/L22}, \href
  {https://ui.adsabs.harvard.edu/abs/2016ApJ...824L..22M} {824, L22}

\bibitem[\protect\citeauthoryear{{Morais} \& {Namouni}}{{Morais} \&
  {Namouni}}{2013}]{MoraisNamouni13b}
{Morais} M.~H.~M.,  {Namouni} F.,  2013, \mn@doi [\mnras]
  {10.1093/mnrasl/slt106}, \href
  {http://adsabs.harvard.edu/abs/2013MNRAS.436L..30M} {436, L30}

\bibitem[\protect\citeauthoryear{{Morais} \& {Namouni}}{{Morais} \&
  {Namouni}}{2016}]{MoraisNamouni16}
{Morais} M.~H.~M.,  {Namouni} F.,  2016, \mn@doi [Celestial Mechanics and
  Dynamical Astronomy] {10.1007/s10569-016-9674-3}, \href
  {http://adsabs.harvard.edu/abs/2016CeMDA.125...91M} {125, 91}

\bibitem[\protect\citeauthoryear{{Morais} \& {Namouni}}{{Morais} \&
  {Namouni}}{2017}]{MoraisNamouni17b}
{Morais} M.~H.~M.,  {Namouni} F.,  2017, \mnras, \href
  {http://doi.org/10.1093/mnrasl/slx125} {472, L1}

\bibitem[\protect\citeauthoryear{{Morbidelli}, {Batygin}, {Brasser}  \&
  {Raymond}}{{Morbidelli} et~al.}{2020}]{mbbr}
{Morbidelli} A.,  {Batygin} K.,  {Brasser} R.,   {Raymond} S.~N.,  2020,
  \mn@doi [\mnras] {10.1093/mnrasl/slaa111}, \href
  {https://ui.adsabs.harvard.edu/abs/2020MNRAS.497L..46M} {497, L46}

\bibitem[\protect\citeauthoryear{{Murray} \& {Dermott}}{{Murray} \&
  {Dermott}}{2000}]{SSDbook}
{Murray} C.~D.,  {Dermott} S.~F.,  2000, {Solar system dynamics}.
{Cambridge University Press}

\bibitem[\protect\citeauthoryear{{Namouni} \& {Morais}}{{Namouni} \&
  {Morais}}{2015}]{NamouniMorais15}
{Namouni} F.,  {Morais} M.~H.~M.,  2015, \mn@doi [\mnras]
  {10.1093/mnras/stu2199}, \href
  {http://adsabs.harvard.edu/abs/2015MNRAS.446.1998N} {446, 1998}

\bibitem[\protect\citeauthoryear{{Namouni} \& {Morais}}{{Namouni} \&
  {Morais}}{2017}]{NamouniMorais17}
{Namouni} F.,  {Morais} M.~H.~M.,  2017, \mn@doi [\mnras]
  {10.1093/mnras/stx290}, 467, 2673

\bibitem[\protect\citeauthoryear{{Namouni} \& {Morais}}{{Namouni} \&
  {Morais}}{2018a}]{NamouniMorais18c}
{Namouni} F.,  {Morais} M.~H.~M.,  2018a, \mn@doi [J. Comp. App. Math.]
  {10.1007/s40314-017-0489-y}, 37, 65

\bibitem[\protect\citeauthoryear{{Namouni} \& {Morais}}{{Namouni} \&
  {Morais}}{2018b}]{NamouniMorais18b}
{Namouni} F.,  {Morais} M.~H.~M.,  2018b, \mn@doi [\mnras]
  {10.1093/mnrasl/sly057}, \href
  {http://adsabs.harvard.edu/abs/2018MNRAS.477L.117N} {477, L117 (Paper I)}

\bibitem[\protect\citeauthoryear{{Namouni} \& {Morais}}{{Namouni} \&
  {Morais}}{2020a}]{NamouniMorais20c}
{Namouni} F.,  {Morais} M. H.~M.,  2020a, arXiv e-prints, \href
  {https://ui.adsabs.harvard.edu/abs/2020arXiv200909773N} {p. arXiv:2009.09773
  (Paper III)}

\bibitem[\protect\citeauthoryear{{Namouni} \& {Morais}}{{Namouni} \&
  {Morais}}{2020b}]{NamouniMorais20d}
{Namouni} F.,  {Morais} M.~H.~M.,  2020b, \mn@doi [\mnras]
  {10.1093/mnras/staa348}, \href
  {https://ui.adsabs.harvard.edu/abs/2020MNRAS.493.2854N} {493, 2854}

\bibitem[\protect\citeauthoryear{{Namouni} \& {Morais}}{{Namouni} \&
  {Morais}}{2020c}]{NamouniMorais20b}
{Namouni} F.,  {Morais} M.~H.~M.,  2020c, \mn@doi [\mnras]
  {10.1093/mnras/staa712}, \href
  {https://ui.adsabs.harvard.edu/abs/2020MNRAS.494.2191N} {494, 2191 (Paper
  II)}

\bibitem[\protect\citeauthoryear{{Nesvorn{\'y}}, {Vokrouhlick{\'y}}, {Dones},
  {Levison}, {Kaib}  \& {Morbidelli}}{{Nesvorn{\'y}} et~al.}{2017}]{Nesvorny17}
{Nesvorn{\'y}} D.,  {Vokrouhlick{\'y}} D.,  {Dones} L.,  {Levison} H.~F.,
  {Kaib} N.,   {Morbidelli} A.,  2017, \mn@doi [\apj]
  {10.3847/1538-4357/aa7cf6}, \href
  {https://ui.adsabs.harvard.edu/abs/2017ApJ...845...27N} {845, 27}

\bibitem[\protect\citeauthoryear{{Oldroyd} \& {Trujillo}}{{Oldroyd} \&
  {Trujillo}}{2021}]{Oldroyd21}
{Oldroyd} W.~J.,  {Trujillo} C.~A.,  2021, \mn@doi [\aj]
  {10.3847/1538-3881/abfb6f}, \href
  {https://ui.adsabs.harvard.edu/abs/2021AJ....162...39O} {162, 39}

\bibitem[\protect\citeauthoryear{{Pfalzner} et~al.,}{{Pfalzner}
  et~al.}{2015}]{Pfalzner15}
{Pfalzner} S.,  et~al., 2015, \mn@doi [\physscr]
  {10.1088/0031-8949/90/6/068001}, \href
  {http://adsabs.harvard.edu/abs/2015PhyS...90f8001P} {90, 068001}

\bibitem[\protect\citeauthoryear{{Portegies Zwart}, {Torres}, {Cai}  \&
  {Brown}}{{Portegies Zwart} et~al.}{2021}]{Portegies21}
{Portegies Zwart} S.,  {Torres} S.,  {Cai} M.~X.,   {Brown} A. G.~A.,  2021,
  \mn@doi [\aap] {10.1051/0004-6361/202040096}, \href
  {https://ui.adsabs.harvard.edu/abs/2021A&A...652A.144P} {652, A144}

\bibitem[\protect\citeauthoryear{{Quinn}, {Tremaine}  \& {Duncan}}{{Quinn}
  et~al.}{1990}]{Quinn90}
{Quinn} T.,  {Tremaine} S.,   {Duncan} M.,  1990, \mn@doi [\apj]
  {10.1086/168800}, \href {http://adsabs.harvard.edu/abs/1990ApJ...355..667Q}
  {355, 667}

\bibitem[\protect\citeauthoryear{{Silsbee} \& {Tremaine}}{{Silsbee} \&
  {Tremaine}}{2018}]{silsbee18}
{Silsbee} K.,  {Tremaine} S.,  2018, \mn@doi [\aj] {10.3847/1538-3881/aaa19b},
  \href {https://ui.adsabs.harvard.edu/abs/2018AJ....155...75S} {155, 75}

\bibitem[\protect\citeauthoryear{{Tiscareno} \& {Malhotra}}{{Tiscareno} \&
  {Malhotra}}{2003}]{TiscarenoMalhotra03}
{Tiscareno} M.~S.,  {Malhotra} R.,  2003, \mn@doi [\aj] {10.1086/379554}, \href
  {http://adsabs.harvard.edu/abs/2003AJ....126.3122T} {126, 3122}

\bibitem[\protect\citeauthoryear{{Tisserand}}{{Tisserand}}{1896}]{Tisserand}
{Tisserand} F.,  1896, {Trait\'e de M\'ecanique C\'eleste. Vol IV}.
{Gauthier-Villards, Paris}

\bibitem[\protect\citeauthoryear{{Trujillo} \& {Sheppard}}{{Trujillo} \&
  {Sheppard}}{2014}]{trujillo14}
{Trujillo} C.~A.,  {Sheppard} S.~S.,  2014, \mn@doi [\nat]
  {10.1038/nature13156}, \href
  {https://ui.adsabs.harvard.edu/abs/2014Natur.507..471T} {507, 471}

\bibitem[\protect\citeauthoryear{{Vokrouhlick{\'y}}, {Nesvorn{\'y}}  \&
  {Dones}}{{Vokrouhlick{\'y}} et~al.}{2019}]{Vokrouhlicky19}
{Vokrouhlick{\'y}} D.,  {Nesvorn{\'y}} D.,   {Dones} L.,  2019, \mn@doi [\aj]
  {10.3847/1538-3881/ab13aa}, \href
  {https://ui.adsabs.harvard.edu/abs/2019AJ....157..181V} {157, 181}

\bibitem[\protect\citeauthoryear{{Volk} \& {Malhotra}}{{Volk} \&
  {Malhotra}}{2013}]{VolkMalhotra13}
{Volk} K.,  {Malhotra} R.,  2013, \mn@doi [\icarus]
  {10.1016/j.icarus.2013.02.016}, \href
  {http://adsabs.harvard.edu/abs/2013Icar..224...66V} {224, 66}

\bibitem[\protect\citeauthoryear{{Volk} \& {Malhotra}}{{Volk} \&
  {Malhotra}}{2017}]{volk17}
{Volk} K.,  {Malhotra} R.,  2017, \mn@doi [\aj] {10.3847/1538-3881/aa79ff},
  \href {https://ui.adsabs.harvard.edu/abs/2017AJ....154...62V} {154, 62}

\makeatother
\end{thebibliography}

\end{document}